\documentclass[%
 reprint,
 floatfix,
 amsmath,amssymb,
 aps, physrev
]{revtex4-2}

\usepackage{graphicx}
\usepackage{dcolumn}
\usepackage{bm}
\usepackage[colorlinks=true, citecolor=blue, linkcolor=blue, urlcolor=blue]{hyperref}
\usepackage{booktabs}
\usepackage{amsmath} 
\usepackage{mathtools}
\usepackage[english]{babel}
\usepackage{orcidlink}

\def\numero{%
  \leavevmode
  \hbox{%
    $\rm N^{\mkern0.8mu\underline{\mkern-0.8mu o\mkern-0.8mu}\mkern0.8mu}$%
  }%
}

\begin{document}

\title{\textit{Ab initio} exploration of hydrogen-intercalated transition metal dichalcogenides: From identifying stable structures to emergent superconductivity}

\author{William Wenborn~\orcidlink{0009-0004-3699-5262}}
\email{william.wenborn@uantwerpen.be}

\author{Jonas Bekaert~\orcidlink{0000-0002-3610-6240}}
\email{jonas.bekaert@uantwerpen.be}
\affiliation{TQC, Department of Physics, University of Antwerp, Universiteitsplein 1, B-2610 Antwerp, Belgium}
\affiliation{COMMIT, Department of Physics, University of Antwerp, Groenenborgerlaan 171, B-2020 Antwerp, Belgium}

\date{\today}

\begin{abstract}
    \noindent {We present a first-principles study of hydrogen-intercalated transition metal dichalcogenides (TMDs), examining the structural, thermodynamic, electronic and superconducting properties of distinct intercalation configurations. In most configurations, the two intercalated hydrogen atoms contribute two electrons per unit cell, creating a band gap and thus rendering the system semiconducting. Among the dynamically stable configurations there is one exception to this rule, with inherently metallic behavior. The pristine phases of these materials are known to exhibit a charge density wave instability, which is fully suppressed by specific hydrogen configurations. Moreover, the semiconducting structures can be metallized by hole doping, albeit that the superconducting critical temperatures ($T_c$) are vanishingly low. On the other hand, electron doping drives the $T_c$ up to 25 K, by populating dense states above the gap. This finding can motivate further experimental exploration of superconductivity in hydrogen intercalated TMDs through ionic-liquid gating or alkali co-intercalation.}
\end{abstract}

\maketitle

\section{Introduction}
Transition-metal dichalcogenides (TMDs) are layered materials with chemical formula MX$_2$, where M is a transition metal (e.g., Ta, Nb, Mo) and X is a chalcogen (S, Se, or Te). Each layer consists of covalently bonded X--M--X atoms, with adjacent layers coupled by van der Waals interactions. TMDs occur in a variety of structural polytypes that differ in both the local coordination of the metal atoms and the stacking sequence of the layers. In particular, the notation 1H denotes a single layer with trigonal-prismatic coordination. In the bulk 2H polytype, these layers are stacked in an AA' sequence, with successive layers rotated by $\pi$~radians relative to each other. This stacking restores inversion symmetry in the bulk, which is absent in the isolated monolayer. Owing to their interplay of electronic, structural, and optical properties, TMDs have become a promising material class for studying correlated quantum phases~\cite{Wilson1975, manzeli20172d, wang2012electronics, Castro2001, Yin2021}. In this work, we focus on the 2H polytypes of TaS$_2$ and NbSe$_2$, two prototypical TMDs in which charge-density-wave (CDW) order coexists with superconductivity~\cite{Wilson1975, manzeli20172d}.

TaS$_2$ and NbSe$_2$ are known to crystallize in several polytypes, including 1T, 2H, 3R, and 6R, whose distinct electronic properties arise from differences in coordination and interlayer stacking~\cite{JELLINEK19629,NAGATA19921259,Wilson1969}. These materials provide a useful platform for studying the interplay between charge-density-wave (CDW) order and superconductivity because both phases are highly sensitive to dimensionality, stacking, and external perturbations. Alternative stacking sequences can significantly modify the balance between CDW order and superconductivity, as demonstrated in 6R-TaS$_2$ and related phase-engineered compounds~\cite{Achari2022,yadav2026}, while external stimuli such as pressure can likewise induce substantial changes in the superconducting state~\cite{Huang2025}. NbSe$_2$ has become a model system for investigating the evolution of intertwined CDW and superconducting orders from the bulk crystal to the monolayer limit~\cite{Ugeda_2015, Zheng2019, Cheung2024}, while comparative studies across the 2H-TMD family have established that the nature of the CDW state depends strongly on dimensionality~\cite{Lin_2020}. Importantly, both 2H-TaS$_2$ and 2H-NbSe$_2$ exhibit CDW ordering vectors along the $\Gamma$--M direction of the Brillouin zone, with $\mathbf{q}_{\mathrm{CDW}}\approx(2/3)\Gamma\mathrm{-M}$ in 2H-TaS$_2$~\cite{Joshi2019} and 2H-NbSe$_2$~\cite{Malliakas2013}. This related ordering vector and the broader similarities in their electronic structures provide a consistent basis for comparing calculations and assessing how external perturbations modify competing ordered phases.\par\smallskip

In bulk 2H-NbSe$_2$, the transition temperatures are $T_{\mathrm{CDW}}=33$~K and $T_c=7.2$~K for superconductivity, whereas bulk 2H-TaS$_2$ has a lower superconducting transition temperature of $T_c\approx0.8$~K~\cite{Wilson1969}. The sensitivity of these competing phases to structural and electronic tuning is evident across different dimensional limits. In monolayer TaS$_2$, $T_c$ increases to $2.2$~K~\cite{navarro-moratalla_enhanced_2016}, while in monolayer NbSe$_2$ the CDW onset temperature rises to ${\sim}145$~K and $T_c$ is reduced to ${\sim}2$~K~\cite{Ugeda_2015}. Beyond dimensionality alone, the phase balance can be manipulated through carrier-density control and chemical modification. Chemical co-intercalation~\cite{zhang2022, Luo2026}, electrostatic gating~\cite{piatti2018, Wu_2022}, and intercalation~\cite{Lian2017, Bekaert2019, Bekaert2022, Petrov2021, Bekaert2024} have all been shown to alter the competition between CDW order and superconductivity, while experimental studies on related TMD systems further highlight the important roles of disorder and electron--phonon coupling in stabilizing these phases~\cite{blagojevic2024}. Together, these studies establish 2H-TaS$_2$ and 2H-NbSe$_2$ as particularly suitable model systems for investigating how intercalation-driven electronic and structural modifications reshape the CDW and superconducting ground states.\par\smallskip

Hydrogen is a particularly attractive intercalant because its small size allows access to coordination environments that are unavailable to heavier atomic species. Hydrogen insertion changes the carrier density through charge transfer to the host layer and also reshapes the local chalcogen coordination around the transition-metal site. In hydride compounds, this combination is associated with record superconducting critical temperatures, such as $T_c=203$~K in H$_3$S~\cite{drozdov2015conventional} and $T_c=250$~K in LaH$_{10}$~\cite{drozdov2019superconductivity,PhysRevLett.122.027001} at pressures of $150$--$200$~GPa. In this work, we propose that layered TMDs provide a complementary route for introducing hydrogen in an ambient-pressure setting, in which H occupies interstitial sites of a structured host.

Among the most recent developments motivating this work are the reports that gate-driven hydrogen intercalation can strongly alter the CDW order in TMDs and induce superconductivity. In particular, protonation of 1T-TiSe$_2$ by ionic-liquid gating was shown to induce superconductivity with $T_c \approx 3.7$~K, while the CDW anomaly in the resistivity is suppressed to about 1\% of its pristine intensity and the CDW onset temperature remains essentially unchanged~\cite{Piatti2023-vg}, with NMR pointing to an inhomogeneous, disorder-driven CDW state in the protonated samples~\cite{Prando_2023}. Notably, the accompanying first-principles calculations indicate that at about one H atom per formula unit, hydrogen acts beyond rigid-band electron doping and partially fills the Ti-$d_{z^2}$ band, mimicking the band filling of superconducting Nb-based 2H TMDs such as 2H-NbSe$_2$~\cite{Piatti2023-vg}. This points to the 2H polytype as a natural host for H-driven superconductivity, further motivating our choice for this material. More generally, the weak van der Waals coupling between adjacent TMD layers provides a natural environment for intercalation, enabling guest species to occupy the interlayer space while preserving much of the host lattice. Likewise, in situ molecular intercalation has very recently been found to enhance $T_c$ in 2H-TaS$_2$ devices~\cite{Pereira_2024}. Together, these advances establish hydrogen-based intercalation as a potential route to engineer the competition between CDW order, semiconducting behavior, and superconductivity in layered TMDs. What remains unresolved, however, is the microscopic structure-property relation in the 2H hosts: which interstitial sites H preferentially occupies in 2H-TaS$_2$ and 2H-NbSe$_2$, which concentrations are thermodynamically and dynamically viable, and how site-selective H incorporation reshapes the electronic structure and superconducting response.\par\smallskip

To address these questions, we first identify and classify the distinct interstitial environments available to hydrogen in the 2H structure, and then determine which of them remain favorable upon relaxation and at increasing H concentration. This establishes the structural and thermodynamic landscape on which the electronic reconstruction must be understood. We next examine how the stable configurations modify the pristine band filling and lattice dynamics, in particular whether hydrogen suppresses the CDW instability by simply shifting the Fermi level into the pre-existing gap or by producing a more substantial reorganization of the host states. Finally, having identified which H arrangements are semiconducting and which retain metallicity, we evaluate their superconducting properties and use carrier doping to assess whether intercalation can provide a route toward enhanced $T_c$ in 2H-TaS$_2$ and 2H-NbSe$_2$.\par\smallskip

\section{Methodology}

All electronic structure calculations were performed using density functional theory (DFT) as implemented in ABINIT~\cite{GONZE2020107042, 10.1063/1.5144261, GONZE2016106} (v.9.10.3). Norm-conserving Perdew-Burke-Ernzerhof (PBE) Goedecker pseudopotentials~\cite{Perdew1996, Goedecker1996, Goedecker1998, vanSetten2018PseudoDojo} were used for Ta~($5d^36s^2$), S~($3s^23p^4$), Nb~($4s^24p^64d^45s^1$), Se~($4s^24p^4$), and H~($1s^1$). Convergence of the cutoff energy was achieved by having total energy differences below $1$~meV per atom and below $0.1$~meV per atom for the \textbf{k}-mesh. This, combined with a potential tolerance of $10^{-12}$, yields an energy cutoff of $50$~Ha and a \textbf{k}-mesh of $16\times16\times6$. Spin-orbit coupling (SOC) is included throughout to account for the relativistic splitting of the Ta and Nb $d$-states. Van der Waals interactions are included via the DFT-D3(BJ) correction~\cite{Grimme2011, Becke2006,10.1063/1.3382344} to capture interlayer cohesion. Wavefunctions were converged when the squared residual over all bands and $k$-points fell below $10^{-24}$, with a potential-iteration tolerance of $10^{-12}$. Structural relaxations were performed using the Broyden–Fletcher–Goldfarb–Shanno (BFGS) algorithm~\cite{broyden1970,fletcher1970,goldfarb1970,shano1970}, and until each Cartesian force component was below $5\times10^{-6}$~Ha/bohr ($0.257$~meV/\AA) and each stress component below $5\times10^{-9}$~Ha/bohr$^3$. 
\par\smallskip

Charge transfer was quantified from the relaxed valence density using Bader partitioning in the atoms-in-molecules framework, evaluated with a grid-based implementation~\cite{doi:10.1021/ar00109a003,Tang_2009}.\par\smallskip

Density functional perturbation theory (DFPT) calculations were also performed within ABINIT~\cite{Baroni2001}. An occupation smearing of $2.5\times10^{-3}$~Ha was determined by comparing phonon frequencies at successive smearing values. Higher smearing artificially removes CDW soft modes~\cite{PhysRevMaterials.8.064803}, so the smearing was reduced until phonon frequencies converged to within $\sim1$~meV. The $\mathbf{q}$-point grid was $8\times8\times3$ with the acoustic sum rule (ASR) enforced. Lastly, wavefunction and potential tolerances were $10^{-24}$ and $10^{-12}$, respectively.\par\smallskip

Superconducting properties were computed using isotropic Eliashberg theory~\cite{Migdal1958, Migdal1960} as implemented in the ANADDB module, with the Eliashberg spectral function defined as
\begin{equation}
\begin{aligned}
    \alpha^2F(\omega) =& \frac{1}{N_F}\sum_{\textbf{k},\textbf{q}}\sum_{m,n,\nu}|g_{\textbf{k},\textbf{k+q},m,n,\nu}|^2\cdot \\
    &\delta (\epsilon_{\textbf{k+q},m}-\epsilon_F)\delta (\epsilon_{\textbf{k},n}-\epsilon_F)\delta (\omega-\omega_{\textbf{q},\nu}).
    \label{eq:Eliashberg_sums}
\end{aligned}
\end{equation}
From this, the total electron-phonon coupling (EPC) constant is obtained as
\begin{equation}
    \lambda_{\text{tot}} = 2\int_0^{\infty}\frac{\alpha^2F(\omega)}{\omega}d\omega.
    \label{eq:EPC_constant}
\end{equation}
Here, $N_F$ is the DOS at $E_F$, $g_{\textbf{k},\textbf{k+q},m,n,\nu}$ are the electron-phonon matrix elements, $\epsilon_{\textbf{k},n}$ are the electronic bands, and $\omega_{\textbf{q},\nu}$ are the phonon eigenvalues.\par\smallskip

$T_c$ is subsequently calculated using the Allen-Dynes modified McMillan formula~\cite{PhysRev.167.331, PhysRevB.12.905}:
\begin{equation}
    T_c=\frac{\hslash \omega_{\text{log}}}{1.2 k_B}\exp{\left( \frac{1.04(1+\lambda_{\text{tot}})}{\mu^*(1+0.62\lambda_{\text{tot}})-\lambda_{\text{tot}}}\right)},
    \label{eq:mcmillan_mod}
\end{equation}
$\omega_{\text{log}}$ is the logarithmic average phonon frequency defined as
\begin{equation}
    \omega_{\text{log}}=\exp{\left( \frac{2}{\lambda_{\text{tot}}}\int_0^{+\infty} \frac{\ln{(\omega) }}{\omega }\alpha^2F(\omega)\,d\omega\right)}.
\end{equation}
Coulomb interactions are not explicitly included in the derivation and are instead represented by a scalar value, the Coulomb pseudopotential ($\mu^*$)~\cite{Morel1962}, which is conventionally set to $0.13$~\cite{PhysRev.167.331}. Equation~\eqref{eq:mcmillan_mod} applies strictly when $\lambda_{\text{tot}} > \mu^*/(1-0.62\mu^*)=0.1414$, i.e., when the denominator is negative. In the weak-coupling limit where $\lambda_{\text{tot}} \leq 0.1414$, the denominator changes sign and $T_c = 0$ by convention.\par\smallskip

\section{Pristine structure}\label{sec:pristine_structure}

\begin{figure*}[ht]
    \centering
    \includegraphics[width=1.0\linewidth]{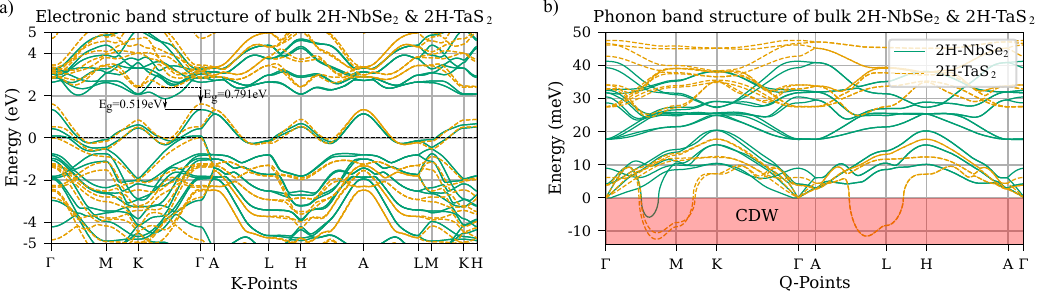}
    \caption{Electronic band structures (a) and phonon band structures (b) of bulk 2H-NbSe$_2$ and bulk 2H-TaS$_2$. An intrinsic band gap (width $\sim 0.79$~eV for TaS$_2$, $\sim 0.52$~eV for NbSe$_2$) resides above $E_F$ and becomes relevant upon hydrogen intercalation. In (b), the imaginary-frequency CDW soft modes are visible near $\mathbf{q}_{\mathrm{CDW}}\approx\tfrac{2}{3}\overline{\Gamma\mathrm{M}}$, consistent with the experimentally observed $3\times3$ charge-density-wave superstructure in both materials.}
    \label{fig:pristine_phonon}
\end{figure*}

Figure~\ref{fig:pristine_phonon}(a) shows the band structures of pristine 2H-TaS$_2$ and NbSe$_2$. Both are metallic with $g(E_F)$ predominantly contributed by transition-metal $d$- and chalcogen $p$-states~\cite{Wilson1969}. An intrinsic gap ($0.791$~eV in TaS$_2$, $0.519$~eV in NbSe$_2$) lies immediately above $E_F$. Figure~\ref{fig:pristine_phonon}(b) shows their phonon dispersions, with CDW soft modes near $\tfrac{2}{3}\overline{\Gamma\mathrm{M}}$, consistent with the established ordering wavevectors in the 2H hosts~\cite{Lin_2020}. TaS$_2$ has 34 valence electrons per cell, with a gap at 36 electrons. Respectively, NbSe$_2$ has 50 valence electrons per cell, with the corresponding gap at 52 electrons. Two intercalated H atoms contribute exactly two electrons per cell, shifting $E_F$ into these gaps and driving the intercalated system semiconducting unless band restructuring prevents this alignment.\par\smallskip

\section{Interstitial positions and thermodynamic stability}
Hydrogen may intercalate into the vdW gap between adjacent X--M--X layers or occupy intralayer sites within the TMD slab. The diversity of accessible positions requires systematic site enumeration and energetic analysis. 

\subsection{Interstitial site determination and classification}
\begin{figure}[ht]
    \centering
    \includegraphics[width=1\linewidth]{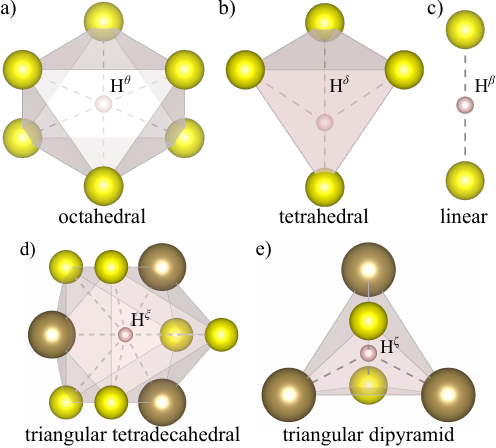}
    \caption{The spatial coordination of each intercalant (pink) position with the associated transition metal (brown) and the chalcogen (yellow). The bridge positions comprise (a)--(c) and the intralayer positions (d) and (e).}
    \label{fig:coordination}
\end{figure}

Possible hydrogen intercalation sites in 2H-TaS$_2$ and 2H-NbSe$_2$ were generated using AFLOW~\cite{Curtarolo_2012}. Symmetry-equivalent positions and sites with unphysically short interatomic separations were removed. After relaxation, only configurations preserving the 2H host structure were retained. Local symmetry breaking induced by H insertion was permitted, provided that the original hexagonal lattice was recovered upon removal of the H atom. A set of recurring, symmetry-equivalent interstitial environments was identified. For clarity, each site was classified according to the polyhedron formed by the nearest-neighbor host atoms and assigned a Greek-letter label corresponding to the number of faces of that polyhedron. This nomenclature is used throughout to distinguish the different interstitial environments. The resulting symmetry-distinct single-H intercalation sites are shown in Fig.~\ref{fig:coordination}. \par\smallskip

The $\theta$ site (a) lies between six S atoms, forming an octahedral-like configuration. The $\delta$ site (b) is tetrahedrally coordinated by S atoms within the van der Waals gap. Depending on the initial structure, the H atom may relax either closer to the three top S atoms (denoted H$^{\delta^*}$) or closer to the bottom S atom (H$^{\delta}$). The $\beta$ site (c) lies slightly offset between two S atoms. These three configurations all lie in the vdW gap between the TaS$_2$ layers and are referred to as `bridge' sites. \par\smallskip

The fourth identified site (d), $\xi$, is located within the TaS$_2$ layer itself directly aligned with Ta atoms, acting as an `intralayer' site embedded in the TMD slab. Lastly, the fifth site (e), $\zeta$, is obtained through a slight vertical shift with respect to $\xi$ and is positioned in between the S atoms. The same configurations apply to 2H-NbSe$_2$, as it has an identical structure to 2H-TaS$_2$. \par\smallskip

These single-H sites were combined to generate double and quadruple intercalation configurations, all subjected to the same relaxation and stability analysis.\par\smallskip

\subsection{Thermodynamic stability}
Thermodynamic stability is assessed here to distinguish favorable intercalation geometries from relaxed but weakly bound arrangements. The same analysis also lets us compare how stability evolves with site type and H concentration on a common footing.\par\smallskip
Thermodynamic stability is quantified by the change in internal energy upon removing $x$ hydrogen atoms, referenced to pristine MX$_2$ (TaS$_2$ or NbSe$_2$) and atomic H:
\begin{equation}
    \Delta U=\frac{x E_{\mathrm{tot}}(\mathrm{H})+E_{\mathrm{tot}}(\mathrm{H}_{N-x}\mathrm{MX}_2)-E_{\mathrm{tot}}(\mathrm{H}_{N}\mathrm{MX}_2)}{x},
\end{equation}
where $x$ represents the number of hydrogen atoms removed from the intercalated system in forming the reference structure. When all hydrogen atoms are removed ($x = N$), this reduces to the binding energy per hydrogen atom:
\begin{equation}
    E_{\mathrm{b}}=\frac{N E_{\mathrm{tot}}(\mathrm{H})+E_{\mathrm{tot}}(\mathrm{MX}_2)-E_{\mathrm{tot}}(\mathrm{H}_{N}\mathrm{MX}_2)}{N}.
    \label{eq:binding_E}
\end{equation}
A positive $E_{\mathrm{b}}$ indicates that the intercalated structure is energetically preferred over its dissociated constituents. The atomic H reference is appropriate since the relevant process is H insertion into pre-existing interstitial sites. $E_{\mathrm{b}}$ relative to molecular hydrogen (H$_2$) follows by subtracting $\sim 1.1$~eV per H atom (the PBE H$_2$ dissociation energy), shifting all values uniformly without affecting the thermodynamic ranking of different configurations. Together, $\Delta U$ and $E_{\mathrm{b}}$ distinguish the stability of individual hydrogen configurations from the overall binding energy of the fully intercalated structure.
\par\smallskip

We also estimate the harmonic zero-point energy ($E_{\text{ZP}}$):
\begin{equation}
    E_{\mathrm{ZP}}=\frac{1}{2}\sum_{\mathbf{q},\nu}\hbar\omega_{\mathbf{q},\nu},
\end{equation}
where the sum runs over all phonon branches $\nu$ and wave vectors $\mathbf{q}$ in the Brillouin zone. This quantity is well defined only for dynamically stable structures, i.e., when all phonon frequencies are real and non-negative throughout the Brillouin zone. In contrast to the total energy, the zero-point energy is an absolute quantity and can be interpreted without taking relative differences. As such, it is most meaningfully computed within the six dynamically stable double-H phases considered here. The resulting values are on the order of $0.55$--$0.72$~eV per cell and are dominated by the H-derived high-frequency modes (see the Supplemental Material~\cite{SM} for specific values). They are much smaller than the static binding energies and therefore do not alter the stability ordering. ZPE differences between the stable phases shift their relative binding energies by at most $\approx 45$~meV per H atom, compared with $E_{\mathrm{b}}$ separations of at least $\approx 0.55$~eV.

\subsubsection{Tantalum Disulfide}

Among bridge sites, H$^{\beta}$ is the most favorable, followed closely by H$^{\delta}$, whereas H$^{\theta}$, although thermodynamically stable, exhibits a substantially lower binding energy, nearly half that of the other two configurations. The spread within this bridge-site family is already substantial, showing that binding is controlled not only by whether H sits in the vdW gap, but also by the chalcogen coordination and by how strongly the host lattice must distort to accommodate that site. Table~\ref{tab:H2_TaS2_energies} summarizes the relaxed structures and $E_{\mathrm{b}}$ per H, with the energies of step-by-step H-removal given in the Supplemental Material~\cite{SM}.\par\smallskip

The weak binding of H$^{\theta}$ is particularly instructive. Although it occupies a highly symmetric octahedral-like cage, the associated contraction of the interlayer spacing shows that this site is stabilized only by pulling the two S planes together, which carries a sizeable structural penalty. By contrast, the smaller separation between H$^{\beta}$ and H$^{\delta}$ indicates that both are competitive bridge configurations, but that their different vertical relaxations still matter at the few-100~meV level. The clear preference for H$^{\delta}$ over H$^{\delta^*}$ further shows that even within the same nominal tetrahedral environment, shifting H toward one chalcogen plane or the other signifies the non-negligible effect of lattice distortion even at low H concentrations.\par\smallskip

The choice of H concentration is also important for the resulting structural and electronic properties. Single and quadruple intercalation do not fully suppress the CDW instability, whereas selected double-intercalation configurations achieve complete suppression while retaining dynamical stability. This makes double intercalation particularly relevant for comparing the relative stability of different H sites in the context of CDW suppression. At higher concentrations, additional H--H interactions can destabilize the structure, as exemplified by H$_2^{\delta}$, which exhibits strong destabilization and dynamical instabilities discussed in the Supplemental Material~\cite{SM}. The strongest binding overall is instead obtained for the intralayer H$^{\xi}$ site ($E_{\mathrm{b}} = 3502$~meV/H), indicating that in TaS$_2$ the energetic gain from coupling H directly to the TaS$_2$ layers outweighs the advantage of remaining in the vdW gap.\par\smallskip
\begin{table}[!t]
\centering
\begin{tabular}{lccccc}
\hline \hline \vspace{1pt}
    Structure & Space group & \multicolumn{1}{c}{$a$ $(\textup{\AA})$} & \multicolumn{1}{c}{$c$ $(\textup{\AA})$} & \multicolumn{1}{c}{$E_{\mathrm{b}}$ (meV)} & \multicolumn{1}{c}{$\Delta q_{\mathrm{H}}$ (e)} \\ \hline 
 H$^{\xi}$TaS$_2$ & $P6_3/mmc$ (\numero 194) & $3.368$ & $12.355$ & $3502$ & $+0.66$ \\
 H$^{\beta}$TaS$_2$ & $Cmcm$ (\numero 63) & $3.374$ & $12.701$ & $3034$ & $-0.22$ \\
 H$^{\delta}$TaS$_2$ & $P6_3mc$ (\numero 186) & $3.392$ & $12.218$ & $2946$ & $-0.16$ \\
 H$^{\theta}$TaS$_2$ & $P6_3/mmc$ (\numero 194) & $3.455$ & $10.412$ & $1663$ & $-0.09$ \\
 H$^{\zeta}$TaS$_2$ & $P6_3/mmc$ (\numero 194) & $3.444$ & $12.051$ & $1535$ & $+0.26$ \\
 \hline \hline
\end{tabular}
\caption{Structural parameters and binding energies per hydrogen atom for double-hydrogen intercalation configurations of 2H-TaS$_2$. $\Delta q_{\mathrm{H}}$ is the mean Bader charge transfer per H atom (negative values indicate H donates electrons).}
\label{tab:H2_TaS2_energies}
\end{table}

\subsubsection{Niobium Diselenide}

Our analogous calculations for 2H-NbSe$_2$ probe which parts of the TaS$_2$ site hierarchy are generic to the 2H host geometry and which conversely depend on the chemistry of the metal-chalcogen layer.\par\smallskip

The same classes of bridge and intralayer sites remain energetically accessible, and H$^{\xi}$ is again the strongest-binding configuration. The changes relative to TaS$_2$, however, are not a simple rigid shift of all binding energies. NbSe$_2$ binds H more weakly at $\beta$, $\delta$, and $\theta$ sites but more strongly at $\xi$ and $\zeta$ (cf.\ Table~\ref{tab:H2_NbSe2_energies} vs.\ Table~\ref{tab:H2_TaS2_energies}), implying that the host dependence is local and site specific rather than a uniform change in H chemical potential. This site sensitivity is most evident at H$^{\delta}$, where the Bader charge reverses sign relative to TaS$_2$. The unusually large $\Delta q_{\mathrm{H}}=+1.16$~e indicates a markedly different local bonding response at the $\delta$ site in NbSe$_2$, consistent with the shorter relaxed Nb--H distance and with the altered stability sequence.\par\smallskip

\begin{table}[htbp]
\centering
\begin{tabular}{lccccc}
\hline \hline \vspace{1pt}
    Structure & Space group & \multicolumn{1}{c}{$a$ $(\textup{\AA})$} & \multicolumn{1}{c}{$c$ $(\textup{\AA})$} & \multicolumn{1}{c}{$E_{\mathrm{b}}$ (meV)} & \multicolumn{1}{c}{$\Delta q_{\mathrm{H}}$ (e)} \\ \hline 
 H$^{\xi}$NbSe$_2$ & $P6_3/mmc$ (\numero 194) & 3.401 & 13.053 & 3801 & +0.43 \\
 H$^{\beta}$NbSe$_2$ & $Cmcm$ (\numero 63) & 3.482 & 13.400 & 2731 & -0.12 \\
 H$^{\delta}$NbSe$_2$ & $P6_3mc$ (\numero 186) & 3.496 & 12.871 & 2623 & +1.16 \\
 H$^{\zeta}$NbSe$_2$ & $P6_3/mmc$ (\numero 194) & 3.478 & 12.717 & 2007 & +0.24 \\
 H$^{\theta}$NbSe$_2$ & $P6_3/mmc$ (\numero 194) & 3.603 & 10.750 & 1613 & +0.00 \\ 
 \hline \hline
\end{tabular}
\caption{Structural parameters and binding energies per hydrogen atom for double-hydrogen intercalation configurations of 2H-NbSe$_2$. $\Delta q_{\mathrm{H}}$ is the mean Bader charge transfer per H atom (negative values indicate H donates electrons).}
\label{tab:H2_NbSe2_energies}
\end{table}

The relaxed $c$ parameter provides a structural counterpart to the binding-energy trends. Most bridge and intralayer sites either expand the vdW gap modestly or leave it close to the intrinsic value, consistent with H occupying the pre-existing free volume between the layers. H$^{\theta}$ forms an exception, as it contracts $c$ to $10.41$~\AA\ in TaS$_2$ and $10.75$~\AA\ in NbSe$_2$, corresponding to reductions of about $15.7\%$ and $17.6\%$, respectively, relative to the corresponding H$^{\xi}$ structures. H$^{\theta}$ bridges two chalcogen planes, forming an octahedral cage. The short H--S distances pull the adjacent layers together, so the contraction constitutes a direct signature of this H configuration. Despite its high coordination, H$^{\theta}$ is relatively weakly bound compared to the other interstitial configurations considered.\par\smallskip

\begin{figure}[htbp]
    \centering
    \includegraphics[width=1\linewidth]{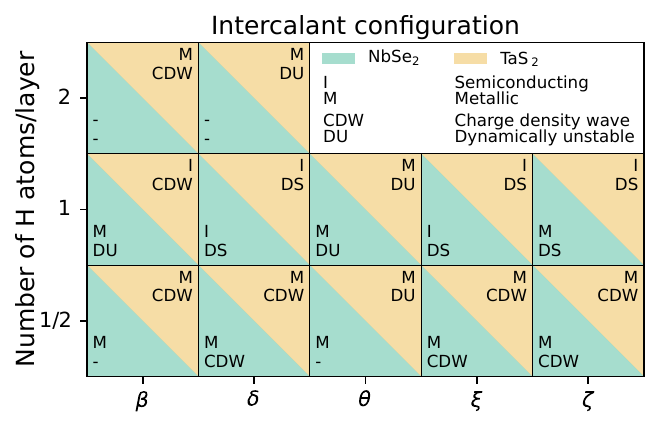}
    \caption{Overview of all intercalation configurations, organized by H content per layer (rows) and interstitial site (columns). Each cell is split diagonally into an NbSe$_2$ half (lower left) and a TaS$_2$ half (upper right), each reporting the electronic character (I, semiconducting; M, metallic) and the dynamical character (DS, dynamically stable; DU, dynamically unstable; CDW, dynamically unstable with the pristine CDW soft mode retained). A dash (--) marks a configuration that was not computed, and empty cells mark structurally inaccessible configurations.}
    \label{fig:matrix_plot}
\end{figure}

Figure~\ref{fig:matrix_plot} provides an overview of all configurations and their dynamical stability and electronic properties. The rows correspond to $1/2$, $1$, and $2$ H atoms per layer, i.e., single, double, and quadruple H intercalation of the 2H unit cell, and the columns to the interstitial sites. Configurations marked with a dash were not computed, since BFGS relaxation already revealed symmetry breaking into a low-symmetry structure for which DFPT becomes prohibitively expensive, so these cases were deprioritized in favor of the higher-symmetry configurations discussed in the main text. The fully empty cells, which occur only for $\theta$, $\xi$, and $\zeta$ at $2$ H atoms per layer, are structurally not accessible, since these sites allow only a single H position per layer and a second H atom therefore cannot be placed there within the 2H unit cell. Additional data on all the structures can be found in the Supplemental Material~\cite{SM}.

\section{Electronic and vibrational properties}
\subsection{\texorpdfstring{TaS$_2$}{TaS2}}

\begin{figure}[htbp]
    \centering
    \includegraphics[width=1\linewidth]{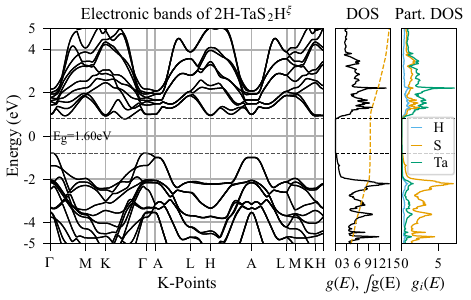}
    \caption{Electronic band structure (left panel), combined density of states (DOS) and total DOS (middle panel), and partial density of states (PDOS, right panel) for the undoped, stable intralayer H$^{\xi}$-intercalated TaS$_2$. The gap opening at $E_F$ illustrates the semiconducting transition induced by the donated electrons through H-intercalation. States above $E_F$ are predominantly Ta-derived, those below are S-dominated.}
    \label{fig:ebands_TaS2xi}
\end{figure}

As established in Sec.~\ref{sec:pristine_structure}, pristine TaS$_2$ lies two electrons short of the intrinsic gap that opens at 36 valence electrons per unit cell. Double intercalation therefore provides a simple electronic starting point, as two H atoms supply exactly the lacking two electrons, so any configuration that does not strongly alter the electronic bands is expected to drive $E_F$ into the gap. Figure~\ref{fig:ebands_TaS2xi} shows this explicitly for the strongest-binding H$^{\xi}$ phase. This filling argument captures the overall behavior of the double-intercalated TaS$_2$ structures.\par\smallskip

The Bader data show highly diverse charge donations, corresponding to different H intercalant sites. Bridge-site H transfers only modest charge to the host ($\Delta q_{\mathrm{H}}$ ranging from $-0.22$ to $-0.09$~e), whereas intralayer H accepts charge ($+0.26$~e for H$^{\zeta}$ and $+0.66$~e for H$^{\xi}$), indicating appreciably different local bonding. Nevertheless, the net redistribution over Ta and S remains consistent with filling the host bands by two electrons, namely Ta loses $2.17$--$2.72$~e/atom and S gains $0.96$--$1.21$~e/atom. The role of site chemistry is therefore to reshape how those electrons are distributed locally, not to overturn the basic gap-filling mechanism.\par\smallskip

This electronic filling is directly tied to the lattice dynamics. Once $E_F$ is pushed into the gap, the metallic screening that supports the pristine CDW soft mode is strongly reduced. That is why H$^{\xi}$-, H$^{\delta}$-, and H$^{\zeta}$-TaS$_2$ are simultaneously gapped and dynamically stable. The same shift of $E_F$ that removes $g(E_F)$ also stiffens the low-energy phonons near $\mathbf{q}_\mathrm{CDW}$. In TaS$_2$, CDW suppression is therefore not a separate consequence of H insertion, but part of the same band-filling mechanism.\par\smallskip

\begin{figure*}[!t]
    \centering
    \includegraphics[width=1\linewidth]{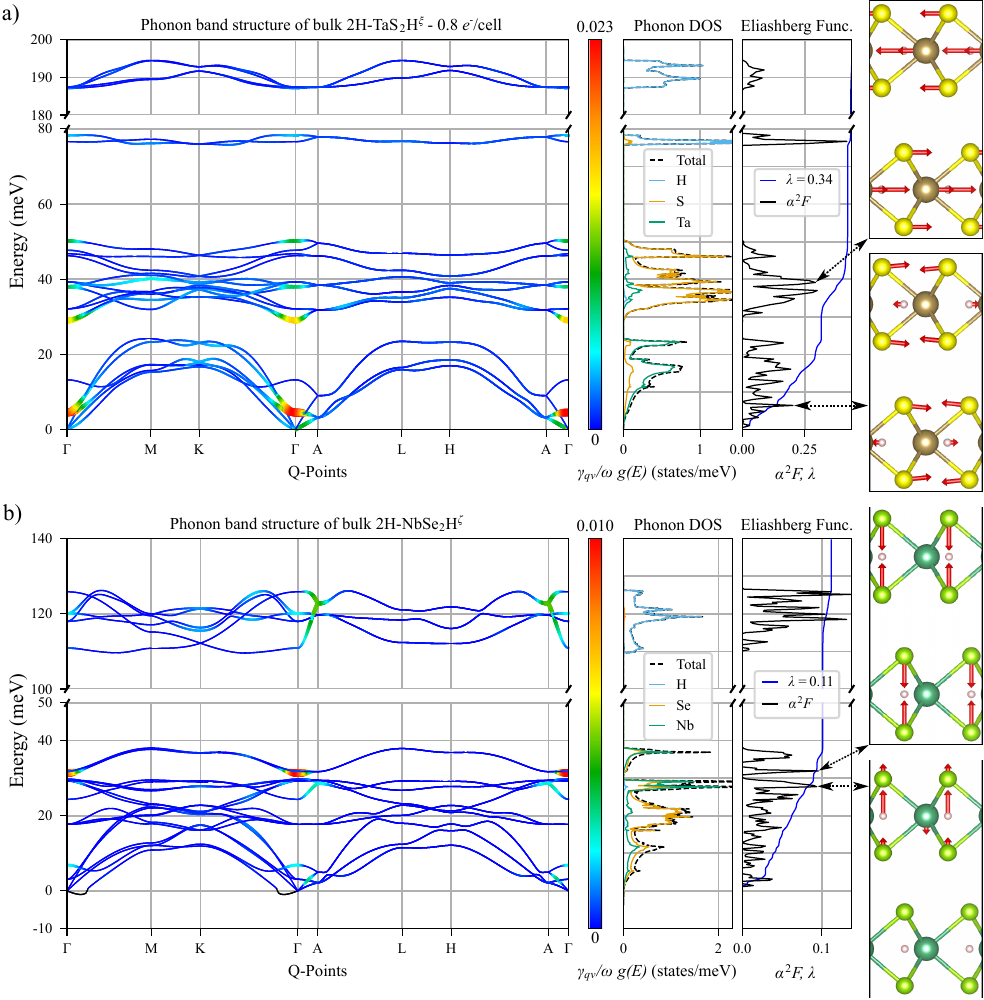}
    \caption{Phonon dispersion and mode-resolved phonon linewidths ($\gamma_{\mathbf{q} \nu}/\omega \propto \alpha^2F$) for, (a), the artificially hole-doped H$^{\xi}$-intercalated TaS$_2$ system ($0.8e^-/\text{cell}$ removed) ensuring induced metallic DOS, and (b), the inherently metallic H$^{\zeta}$-intercalated NbSe$_2$. The phonon dispersion plots reveal the elimination of CDW instabilities, while the mode-resolved linewidths reveal contributions to the total EPC ($\lambda_{\text{tot}}$). Both structures show contributions to the PhDOS from hydrogen at high frequencies, leading to a reduced contribution to the spectral function. To visually convey the atomic behavior, the right-hand panel illustrates the atomic displacements corresponding to the most prominent peaks measured in the Eliashberg spectral function $\alpha^2F$.}
    \label{fig:phonon_dispersion}
\end{figure*}

Mechanical routes to metallization are not effective for this phase, as neither $c$-axis pressure up to $10$~GPa nor in-plane biaxial strain closes the H$^{\xi}$-TaS$_2$ gap (see Supplemental Material~\cite{SM}), motivating a charge-doping approach instead. To test whether restoring charge carrier levels alone is enough to recover superconductivity, we hole-dope H$^{\xi}$-TaS$_2$ by removing $0.8e^-$/cell, reducing the valence electron count to $35.2~e^-$. This shifts $E_F$ back into the valence bands without changing the relaxed H geometry, restoring a finite $g(E_F)$. The resulting phonon dispersion [Fig.~\ref{fig:phonon_dispersion}(a)] shows that, despite this restored metallicity, the pristine CDW instability does not reappear. The H-induced relaxation of the host lattice around the intralayer H$^{\xi}$ site alters the orbital character of the repopulated valence bands relative to pristine TaS$_2$, so the Fermi-surface nesting responsible for the pristine CDW phonon anomaly is not reproduced merely by restoring the electron count. Metallicity is therefore necessary but not sufficient to revive the CDW instability. Self-consistent DFPT for this hole-doped state then gives only $\lambda_{\text{tot}} \approx 0.337$ and $\omega_{\text{log}} \approx 196$~K, yielding $T_c \approx 0.071$~K. Suppressing the CDW by doping does not uncover a stronger superconducting state for the case of exactly two H atoms per cell. It instead produces a metal with only modest EPC once carriers are reintroduced. \par\smallskip 

The sensitivity of this result to the Coulomb pseudopotential was also tested. Varying $\mu^*$ between $0.10$ and $0.16$ changes $T_c$ from $0.26$~K to $0.01$~K (a $\sim 96\%$ spread), which is a large relative change but leaves $T_c$ negligible in absolute terms.\par\smallskip

To further identify which vibrational modes drive the superconducting pairing in this metallic state, we evaluated the isotope effect by varying the atomic masses of H, S, and Ta independently while the electronic structure and force constants remain fixed~\cite{Maxwell1950,BCS1957} (full details in the Supplemental Material~\cite{SM}). Fitting $T_c(M)$ to $T_c \propto M^{-\alpha}$ yields a very small hydrogen exponent, $\alpha_{\mathrm{H}} \approx 0.016$, contrasted with substantially larger values for the host-lattice atoms, namely $\alpha_{\mathrm{S}} \approx 0.297$ and $\alpha_{\mathrm{Ta}} \approx 0.327$. This demonstrates that H-derived vibrations contribute only weakly to the total pairing interaction, and that the dominant electron-phonon coupling originates from Ta and S phonon modes, consistent with the high-frequency H optical branches carrying negligible weight in $\alpha^2F(\omega)$. This is a general trend for H-intercalated TaS$_2$ rather than a universal limitation of hydrogen as an intercalant. In host materials such as hydrogenated monolayer MgB$_2$, strong vibrational hybridization between the H and host-lattice modes couples the H vibrations directly to the electronic states at $E_F$ and substantially boosts $T_c$~\cite{Bekaert2019}. Here, by contrast, the H-derived modes remain largely isolated in frequency from the Ta- and S-dominated branches. Furthermore, hydrogen has a negligible contribution to the lower lying Ta- and S-bands, as seen in the phonon DOS of Fig.~\ref{fig:phonon_dispersion}(a). This limits the hybridization and hence hydrogen's direct contribution to the pairing.\par\smallskip

Beyond this single self-consistently doped case, we extend the doping analysis using the rigid-band approximation. Here, the phonon frequencies, eigenvectors, and electron-phonon matrix elements of the neutral, semiconducting H$^{\xi}$-TaS$_2$ structure are held fixed, and only $E_F$ is shifted through this frozen electronic structure to emulate further hole or electron doping, with $N_F$ and the Eliashberg sums in Eq.~\eqref{eq:Eliashberg_sums} re-evaluated at each shifted filling. We adopt this approach because direct DFPT calculations for the electron-doped cases did not converge within our workflow. The drawback is that it cannot capture any doping-induced change to the lattice geometry or the electron-phonon matrix elements themselves, unlike the single self-consistently doped point discussed above. With this caveat in mind, the rigid-band trends clarify the asymmetry between hole and electron doping. Hole doping re-enters comparatively broad valence states and leaves $T_c$ negligible ($\sim 0$~K for $-0.8e^-/$cell, and $0.27$~K at $-1.0e^-/$cell). Electron doping, by contrast, pushes $E_F$ into sharply peaked states above the gap, driving a rapid increase in the estimated $T_c$ to $6.08$~K at $+0.2e^-/$cell, $13.60$~K at $+0.6e^-/$cell, and $25.10$~K at $+1.0e^-/$cell. Because these rigid-band values retain the phonon spectral function of the undoped structure rather than one relaxed and recomputed self-consistently at each filling, they should be read as upper-bound trend indicators rather than quantitative predictions. Even with that caveat, the result identifies the conduction band, rather than the hole-doped valence bands, as the promising direction for superconductivity enhancement in intercalated TaS$_2$.\par\smallskip

\subsection{\texorpdfstring{NbSe$_2$}{NbSe2}}

NbSe$_2$ follows the same basic filling logic as TaS$_2$, but with a narrower margin between semiconducting and semimetallic behavior. The pristine gap occurs at 52 electrons per cell, so double intercalation again tends to place $E_F$ in the gap. H$^{\xi}$- and H$^{\delta}$-NbSe$_2$ do so cleanly, whereas H$^{\zeta}$ leaves a small band overlap instead of opening a full gap. The dominant effect of two intercalated H atoms is therefore again band filling of the host structure rather than the creation of an entirely new band structure due to fundamental changes caused by intercalation.\par\smallskip

The charge transfer, however, is more site-sensitive than in TaS$_2$. Table~\ref{tab:H2_NbSe2_energies} gives a mean $\Delta q_{\mathrm{H}}\approx +0.34$~e per H, with H$^{\zeta}$ slightly smaller at $+0.24$~e. The clearest outlier is H$^{\delta}$-NbSe$_2$, where $\Delta q_{\mathrm{H}} = +1.16$~e, opposite in sign to H$^{\delta}$-TaS$_2$ ($-0.16$~e). This reversal indicates that the local H--host interaction at the $\delta$ site changes qualitatively between the two materials, even though the overall electronic outcome remains similar.\par\smallskip

The vibrational response mirrors the electronic one. The configurations in which the gap opens are also those in which the CDW soft mode is most strongly suppressed. For the three dynamically stable NbSe$_2$ phases, the remaining imaginary frequencies near $\Gamma$ are of the order of $1$~meV and are only weakly reduced by a larger plane-wave cutoff (Supplemental Material~\cite{SM}). We therefore treat them as numerical remnants of the dynamical-matrix calculation rather than physical instabilities. The ASR-enforced interpolation and the exclusion of these sub-meV modes from $\alpha^2F(\omega)$ and $\lambda_{\text{tot}}$ introduce only a small quantitative uncertainty, and, as will be discussed in the following section, do not significantly affect the results.\par\smallskip

\begin{figure}[htbp]
    \centering
    \includegraphics[width=1\linewidth]{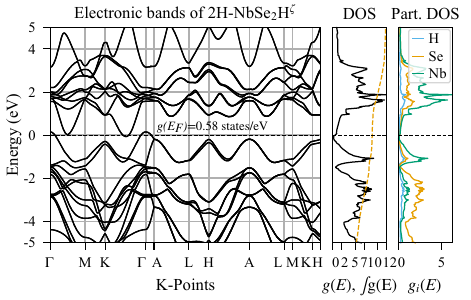}
    \caption{Electronic band structure (left panel), combined density of states (DOS) and total DOS (middle panel), and partial density of states (PDOS, right panel) for the stable intralayer H$^{\zeta}$-intercalated NbSe$_2$. The Fermi level overlaps a conduction band near $\tfrac{1}{2}(M$--$\Gamma)$ and the valence band at $\Gamma$, preserving semimetallic behavior.}
    \label{fig:ebands_NbSe2zeta}
\end{figure}

H$^{\zeta}$-NbSe$_2$ is the key exception because the band edges fail to separate completely. As shown in Fig.~\ref{fig:ebands_NbSe2zeta}, $E_F$ overlaps a conduction band near $\tfrac{1}{2}(M$--$\Gamma)$ and the valence band at $\Gamma$, leaving a small semimetallic DOS of $g(E_F)=0.58$~states/eV. This makes H$^{\zeta}$-NbSe$_2$ a useful test of whether merely preserving metallicity is enough to sustain appreciable superconductivity after intercalation.\par\smallskip

The finite DOS yields only $\lambda_{\text{tot}} \approx 0.11$ and $\omega_{\text{log}} \approx 140$~K. Since $\lambda_{\text{tot}}<\mu^*/(1-0.62\mu^*)=0.1414$, the Allen-Dynes denominator changes sign and $T_c$ is taken to be zero. The mode-resolved linewidths in Fig.~\ref{fig:phonon_dispersion} show that the leading EPC contributions come from layer-breathing modes and H oscillations against individual layers. H-dominated modes near $120$~meV remain visible in $\alpha^2F$, but their high frequency keeps their contribution to $\lambda_{\text{tot}}$ small. Intercalation can therefore preserve a residual metallic state in NbSe$_2$ without recovering the low-frequency coupling strength required for superconductivity. \par\smallskip

\section{Conclusion}

We have presented a first-principles study of hydrogen-intercalated 2H-TaS$_2$ and 2H-NbSe$_2$, investigating the structural, electronic, and superconducting consequences of H insertion. Five distinct sites were identified ($\delta$, $\beta$, $\theta$, $\xi$, $\zeta$), spanning interlayer bridge and intralayer coordination environments. Double intercalation (one H per layer) suppresses the CDW instability, yielding dynamically stable structures.\par\smallskip

The two electrons contributed per unit cell shift $E_F$ into the intrinsic gap of the pristine host, rendering all configurations semiconducting except H$^{\zeta}$-NbSe$_2$. Self-consistent DFPT for hole-doped H$^{\xi}$-TaS$_2$ restores metallicity but merely yields $T_c = 0.071$~K ($\lambda_{\text{tot}}=0.337$, $\omega_{\text{log}}=196$~K), confirming that H insertion alone does not enhance superconductivity. Isotope-mass analysis of this metallic state further shows that the pairing is dominated by Ta and S modes ($\alpha_{\mathrm{Ta}} \approx 0.327$, $\alpha_{\mathrm{S}} \approx 0.297$), while the H contribution is negligible ($\alpha_{\mathrm{H}} \approx 0.016$).\par\smallskip

Rigid-band electron doping shows a monotonic $T_c$ increase reaching $\sim25$~K at $+1.0e^-/$cell, as $E_F$ enters sharply peaked states above the gap. Experimental access to these regimes could be provided by ionic-liquid gating or alkali co-intercalation, both of which can tune carrier density while preserving the 2H configuration. The analogous behavior of TaS$_2$ and NbSe$_2$ supports a common 2H-TMD mechanism, suggesting analogous gap-mediated intercalation effects in other members of this family.\par\smallskip

\section*{Acknowledgments}
The authors acknowledge the use of computational resources provided by the VSC (Flemish Supercomputer Center), funded by the Research Foundation - Flanders (FWO) and the Flemish Government. J. B. acknowledges support as a Senior Postdoctoral Fellow of Research Foundation-Flanders (FWO) under Fellowship No.\ 12ZZ323N.

\bibliography{references}

\end{document}


\title{Supplemental Material: \textit{Ab initio} exploration of hydrogen-intercalated transition metal dichalcogenides: From identifying stable structures to emergent superconductivity}
\author{William Wenborn}
\author{Jonas Bekaert}
\maketitle

\section{Thermodynamic Analysis of Intercalation}

Step-by-step hydrogen removal energies are compiled here to support the binding-energy comparisons in the main text. Reporting the sequential removal energies, rather than only the total $E_{\mathrm{b}}$, makes clear whether adding the second or fourth H atom remains favorable once interlayer coupling or direct H--H repulsion is included. 

The superscript labels the interstitial site and the subscript gives the number of H atoms per layer (omitted when equal to one). The 2H cell has two layers, so $\mathrm{H}^{\xi}_{1/2}\mathrm{TaS}_2$ contains one H atom per cell, $\mathrm{H}^{\xi}\mathrm{TaS}_2$ two (one per layer, as in Tables~I and~II of the main text), and $\mathrm{H}^{\xi}_{2}\mathrm{TaS}_2$ four. $N_H$ is the number of H atoms per cell.

$\Delta U_{n\rightarrow n-1}$ denotes the energy to remove one H atom from the $n$-H intercalated structure:
\begin{align}
    \Delta U_{n \rightarrow n-1} = E(\mathrm{H}_{n-1}\mathrm{TaS}_2) + E(\mathrm{H}) - E(\mathrm{H}_n\mathrm{TaS}_2),
\end{align}
with the final step $\Delta U_{1 \rightarrow 0} = E(\mathrm{TaS}_2) + E(\mathrm{H}) - E(\mathrm{H}_{1/2}\mathrm{TaS}_2)$. As a concrete example for the highest-binding intralayer site H$^{\xi}$, the removal of the first and second H atoms yields:
\begin{align}
    \mathrm{H}^{\xi}\mathrm{TaS}_2 &\xrightarrow[]{-\mathrm{H}^{\xi}} \mathrm{H} + \mathrm{H}_{1/2}^{\xi}\mathrm{TaS}_2 & \Delta U_1 &= 3522~\mathrm{meV}, \\
    \mathrm{H}_{1/2}^{\xi}\mathrm{TaS}_2 &\xrightarrow[]{-\mathrm{H}^{\xi}} \mathrm{H} + \mathrm{TaS}_2 & \Delta U_2 &= 3481~\mathrm{meV},
\end{align}
demonstrating that successive removals are inequivalent even at the same crystallographic site, reflecting interlayer coupling beyond the nearest-layer vdW interaction. Because the two H atoms occupy symmetry-equivalent sites in different layers, the small difference between $\Delta U_1$ and $\Delta U_2$ isolates that residual layer-to-layer coupling rather than a change of local site type.

For the $\beta$-site configurations:
\begin{align}
    \mathrm{H}^{\beta}\mathrm{TaS}_2 & \xrightarrow[]{-\mathrm{H}^{\beta}} \mathrm{H} + \mathrm{H}_{1/2}^{\beta}\mathrm{TaS}_2 & \Delta U &= 2962~\mathrm{meV}, \\
    \mathrm{H}_{1/2}^{\beta}\mathrm{TaS}_2 & \xrightarrow[]{-\mathrm{H}^{\beta}} \mathrm{H} + \mathrm{TaS}_2 & \Delta U &= 3105~\mathrm{meV} .\label{eq:Hbeta_bind}
\end{align}
The $\beta$ site accommodates only one H per layer, both H atoms occupy near-identical positions, and the small difference in sequential removal energies ($2962$ vs.\ $3105$~meV) reflects interlayer vdW modulation. The same pattern holds for $\theta$:
\begin{align}
    \mathrm{H}^{\theta}\mathrm{TaS}_2 & \xrightarrow[]{-\mathrm{H}^{\theta}} \mathrm{H} + \mathrm{H}_{1/2}^{\theta}\mathrm{TaS}_2 & \Delta U &= 1604~\mathrm{meV}, \\
    \mathrm{H}_{1/2}^{\theta}\mathrm{TaS}_2 & \xrightarrow[]{-\mathrm{H}^{\theta}} \mathrm{H} + \mathrm{TaS}_2 & \Delta U &= 1722~\mathrm{meV} .\label{eq:Htheta_bind}
\end{align}

At quadruple concentration, H--H repulsion significantly reduces $\Delta U$ for the $\delta$ site:
\begin{align}
    \mathrm{H}_2^{\delta}\mathrm{TaS}_2 & \xrightarrow[]{-2\mathrm{H}^{\delta}} 2\mathrm{H} + \mathrm{H}^{\delta}\mathrm{TaS}_2 & \Delta U &= 1735~\mathrm{meV}, \\
    \mathrm{H}^{\delta}\mathrm{TaS}_2 & \xrightarrow[]{-\mathrm{H}^{\delta}} \mathrm{H} + \mathrm{H}_{1/2}^{\delta}\mathrm{TaS}_2 & \Delta U &= 2912~\mathrm{meV}, \\
    \mathrm{H}_{1/2}^{\delta}\mathrm{TaS}_2 & \xrightarrow[]{-\mathrm{H}^{\delta}} \mathrm{H} + \mathrm{TaS}_2 & \Delta U &= 2979~\mathrm{meV} .\label{eq:Hdelta_bind}
\end{align}
The reduction from double to quadruple intercalation (from $\sim$2950 to 1735~meV/H) is dominated by direct H--H repulsion at high concentration. For H$^{\delta}_2$, $E_{\mathrm{b}}$ drops by $\sim 0.61$~eV per H relative to H$^{\delta}$ (2341 vs 2946~meV/H). The alternative H$^{\delta*}_2$ variant is further destabilized (1651~meV/H), consistent with stronger repulsive overlap in its relaxed geometry. This reduced binding energy mirrors the dynamical instability of H$^{\delta}_2$ in the phonon analysis of the main text.
The comparison separates two effects: the mild layer-to-layer modulation present at $N_H=2$ and the much stronger crowding penalty that appears once both layers are heavily populated.

Comparing H$_2^{\delta}$ and H$_2^{\delta^*}$ directly confirms the unstarred geometry as more stable:
\begin{align}
    \mathrm{H}_2^{\delta}\mathrm{TaS}_2 & \xrightarrow[]{-4\mathrm{H}^{\delta}} 4\mathrm{H} + \mathrm{TaS}_2 & \Delta U &= 2341~\mathrm{meV}, \\
    \mathrm{H}_2^{\delta^*}\mathrm{TaS}_2 & \xrightarrow[]{-4\mathrm{H}^{\delta^*}} 4\mathrm{H} + \mathrm{TaS}_2 & \Delta U &= 1651~\mathrm{meV}.
    \label{eq:Hdelta*_bind}
\end{align}

Finally, for the highest-binding intralayer site H$^{\xi}$:
\begin{align}
    \mathrm{H}^{\xi}\mathrm{TaS}_2 & \xrightarrow[]{-\mathrm{H}^{\xi}} \mathrm{H} + \mathrm{H}_{1/2}^{\xi}\mathrm{TaS}_2 & \Delta U &= 3522~\mathrm{meV}, \\
    \mathrm{H}_{1/2}^{\xi}\mathrm{TaS}_2 & \xrightarrow[]{-\mathrm{H}^{\xi}} \mathrm{H} + \mathrm{TaS}_2 & \Delta U &= 3481~\mathrm{meV} .\label{eq:Hxi_bind}
\end{align}

\begin{table*}[t]
    \caption{Binding energies for configurations not listed in the main-text Tables I/II, except the entry H$^{\delta}$ ($E_{\rm cut}=80$~Ha), which repeats the main-text H$^{\delta}$-TaS$_2$ structure at a higher cutoff as a convergence test (Sec.~\ref{sec:convergence}). For each structure we list the number of H atoms $N_H$, space-group number, relaxed lattice parameters, and binding energy per H atom.}
    \label{tab:supp_binding}
    \begin{ruledtabular}
    \begin{tabular}{lccccc}
        Configuration & $N_H$ & Space group & $a$ (\AA) & $c$ (\AA) & $E_{\mathrm{b}}$ (meV/H) \\ \hline
        \multicolumn{6}{l}{TaS$_2$} \\
        H$_{1/2}^{\xi}$-TaS$_2$ & 1 & \numero 187 & 3.358 & 12.111 & 3481 \\
        H$_{1/2}^{\beta}$-TaS$_2$ & 1 & \numero 12 & 3.370 & 12.322 & 3105 \\
        H$_{1/2}^{\delta}$-TaS$_2$ & 1 & \numero 156 & 3.373 & 11.948 & 2979 \\
        H$^{\delta}$ ($E_{\text{cut}}=80$~Ha)-TaS$_2$ & 2 & \numero 186 & 3.393 & 12.218 & 2958 \\
        H$_{1/2}^{\delta}$H$_{1/2}^{\theta}$TaS$_2$ & 2 & \numero 156 & 3.412 & 11.957 & 2730 \\
        H$_{1/2}^{\theta}$H$_{1/2}^{\beta}$-TaS$_2$ & 2 & \numero 63 & 3.314 & 12.098 & 2355 \\
        H$_2^{\delta}$-TaS$_2$ & 4 & \numero 194 & 3.378 & 13.907 & 2341 \\
        H$^{\delta}$H$^{\delta*}$-TaS$_2$ & 4 & \numero 186 & 3.378 & 13.907 & 2341 \\
        H$^{\theta}$H$^{\delta}$-TaS$_2$ & 4 & \numero 186 & 3.321 & 13.226 & 1991 \\
        H$^{\theta}$H$^{\beta}$-TaS$_2$ & 4 & \numero 63 & 3.316 & 13.369 & 1927 \\
        H$^{\xi}$H$^{\zeta}$-TaS$_2$ & 4 & \numero 194 & 3.491 & 12.148 & 1910 \\
        H$_{1/2}^{\theta}$-TaS$_2$ & 1 & \numero 164 & 3.392 & 11.198 & 1722 \\
        H$_2^{\delta*}$-TaS$_2$ & 4 & \numero 194 & 3.135 & 13.642 & 1651 \\
        H$_{1/2}^{\zeta}$-TaS$_2$ & 1 & \numero 187 & 3.400 & 11.830 & 1560 \\
        \hline
        \multicolumn{6}{l}{NbSe$_2$} \\
        H$_{1/2}^{\xi}$-NbSe$_2$ & 1 & \numero 187 & 3.415 & 12.588 & 3773 \\
        H$_{1/2}^{\theta}$H$_{1/2}^{\xi}$-NbSe$_2$ & 2 & \numero 156 & 3.506 & 12.570 & 3222 \\
        H$_{1/2}^{\zeta}$H$_{1/2}^{\xi}$-NbSe$_2$ & 2 & \numero 187 & 3.439 & 12.870 & 2890 \\
        H$_{1/2}^{\delta}$H$_{1/2}^{\xi}$-NbSe$_2$ & 2 & \numero 156 & 3.430 & 12.816 & 2745 \\
        H$_{1/2}^{\beta}$H$_{1/2}^{\delta}$-NbSe$_2$ & 2 & \numero 194 & 3.483 & 13.401 & 2731 \\
        H$_{1/2}^{\delta}$H$_{1/2}^{\theta}$-NbSe$_2$ & 2 & \numero 164 & 3.523 & 12.527 & 2654 \\
        H$_{1/2}^{\delta}$-NbSe$_2$ & 1 & \numero 156 & 3.464 & 12.531 & 2650 \\
        H$^{\xi}$H$^{\zeta}$-NbSe$_2$ & 4 & \numero 194 & 3.401 & 13.055 & 2444 \\
        H$_{1/2}^{\zeta}$-NbSe$_2$ & 1 & \numero 187 & 3.457 & 12.401 & 2060 \\
        H$_{1/2}^{\theta}$-NbSe$_2$ & 1 & \numero 164 & 3.497 & 11.646 & 1619 \\
    \end{tabular}
    \end{ruledtabular}
\end{table*}

The table shows that the energetically relevant comparison is not simply between different H counts, but between different ways of distributing H across the two layers. Single-H structures quantify the local binding of an isolated intercalant, mixed double-H structures show how that binding changes once both layers are occupied, and quadruple-H structures mark the onset of strong H--H crowding. The double-H window remains the most relevant regime for the stable phases discussed in the main text because it combines strong binding with the absence of the large repulsive penalties seen at $N_H=4$.

For the six dynamically stable double-H phases, we also evaluated the harmonic zero-point energy (ZPE) in order to estimate the vibrational correction associated with the light H atoms. The ZPE ranges from $549.62$ to $718.74$~meV per cell. TaS$_2$ systematically exhibits a larger ZPE than NbSe$_2$, consistent with slightly stiffer H-dominated vibrational modes in the Ta host. Within TaS$_2$, H$^{\xi}$ and H$^{\zeta}$ are nearly the same, with ZPEs of $718.74$ and $716.64$~meV per cell, respectively, while H$^{\delta}$ is lower at $675.38$~meV per cell. In NbSe$_2$, H$^{\xi}$ again yields the largest ZPE, $638.54$~meV per cell, whereas H$^{\zeta}$ and H$^{\delta}$ are very similar, with values of $549.62$ and $552.60$~meV per cell. These vibrational corrections are small compared with the static binding energies and therefore do not alter the energetic ordering discussed in the main text. A decomposition of the DOS-based ZPE shows that the dominant contribution arises from H, while the transition-metal and chalcogen contributions are substantially smaller, as expected from the large weight of high-frequency H-derived vibrations. We emphasize, however, that these atom-resolved values should be interpreted only as a projection-based decomposition of the total harmonic ZPE, not as uniquely defined or independently measurable zero-point energies of the individual atomic species.

\begin{table*}[htbp]
    \caption{Zero-point energies computed for the stable structures and the relative contributions from the H, M, and X atoms, where M is the transition metal and X is the chalcogen. All values were computed using strictly positive phonon frequencies. The ZPEs are small relative to the binding energies, so they do not change the stability ordering of the structures.}
    \label{tab:suppl_zpe}
    \begin{ruledtabular}
        \begin{tabular}{lcccc}
            Configuration & $E_{\text{ZP,H}}$ & $E_{\text{ZP,M}}$& $E_{{\text{ZP},X}}$ & $E_{\text{ZP}}$ (meV/cell) \\ \hline
            H$^{\xi}$-TaS$_2$ & 445.97 & 58.70 & 214.07 & 718.74 \\
            H$^{\zeta}$-TaS$_2$ & 443.79 & 58.79 & 214.08 & 716.64 \\
            H$^{\delta}$-TaS$_2$ & 389.74	& 54.34 & 231.31 & 675.38 \\ \hline
            H$^{\xi}$-NbSe$_2$ & 416.58 & 89.03 & 132.94 & 638.54 \\
            H$^{\zeta}$-NbSe$_2$ & 347.57 & 75.98 & 126.08 & 549.62 \\
            H$^{\delta}$-NbSe$_2$ & 325.78 & 83.84 & 142.98 & 552.60 \\
        \end{tabular}
    \end{ruledtabular}
\end{table*}

\section{Electronic band structures}

Band structures for configurations not shown in the main text are collected here, grouped by material.

\subsection{\texorpdfstring{TaS$_2$}{TaS2}}

\begin{figure}[H]
    \begin{minipage}{0.48\textwidth}
        \includegraphics[width=\linewidth]{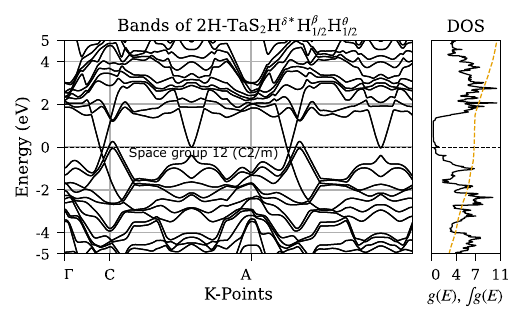}
        \caption{Band structure of 2H-TaS$_2$--H$_2^{\delta^*/\beta/\theta}$. Metallic ($g(E_F)=0.87$~states/eV). Note the strong change in space group.}
        \label{fig:suppl_tas2_hxi}
    \end{minipage}\hfill
    \begin{minipage}{0.48\textwidth}
        \includegraphics[width=\linewidth]{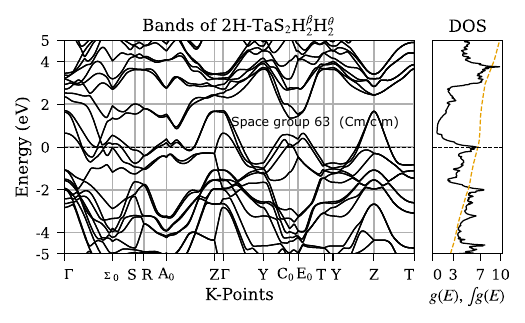}
        \caption{Band structure of 2H-TaS$_2$--H$_2^{\beta/\theta}$. Metallic ($g(E_F)=6.41$~states/eV). The high DOS reflects overlapping H-derived states.}
        \label{fig:suppl_tas2_h4beta_theta}
    \end{minipage}
\end{figure}

\begin{figure}[H]
    \begin{minipage}{0.48\textwidth}
        \includegraphics[width=\linewidth]{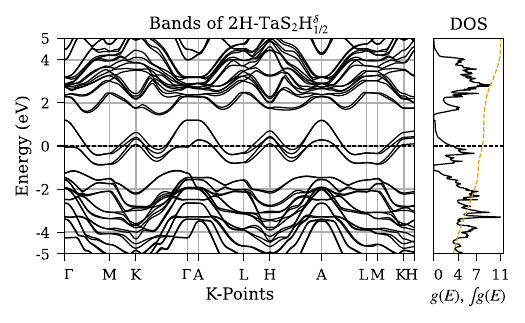}
        \caption{Band structure of 2H-TaS$_2$--H$_{1/2}^{\delta}$. Metallic ($g(E_F)=3.03$~states/eV). The valence edge is pinned just below $E_F$.}
        \label{fig:suppl_tas2_hdelta}
    \end{minipage}\hfill
    \begin{minipage}{0.48\textwidth}
        \includegraphics[width=\linewidth]{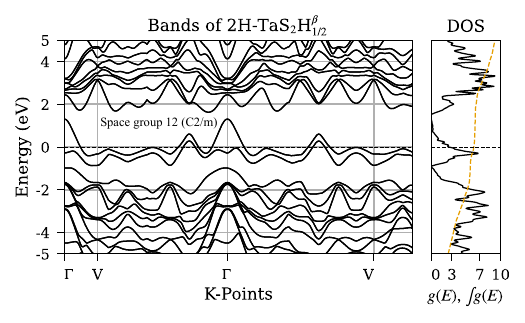}
        \caption{Band structure of 2H-TaS$_2$--H$_{1/2}^{\beta}$. Metallic ($g(E_F)=2.73$~states/eV). Note the strong change in space group.}
        \label{fig:suppl_tas2_hbeta}
    \end{minipage}
\end{figure}

\begin{figure}[H]
    \begin{minipage}{0.48\textwidth}
        \includegraphics[width=\linewidth]{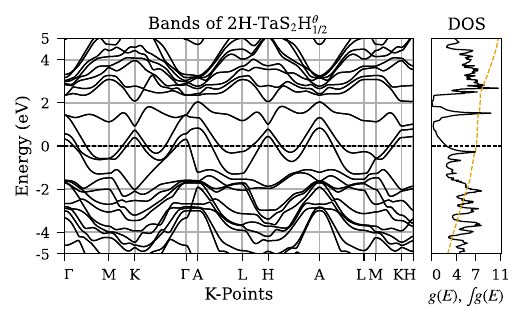}
        \caption{Band structure of 2H-TaS$_2$--H$_{1/2}^{\theta}$. Metallic ($g(E_F)=2.32$~states/eV). The bands are weakly perturbed relative to pristine TaS$_2$.}
        \label{fig:suppl_tas2_htheta}
    \end{minipage}\hfill
    \begin{minipage}{0.48\textwidth}
        \includegraphics[width=\linewidth]{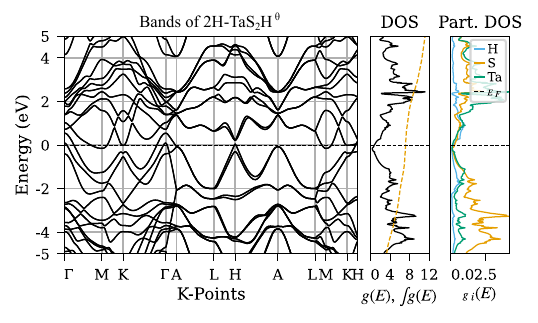}
        \caption{Band structure of 2H-TaS$_2$--H$^{\theta}$. Metallic ($g(E_F)=0.85$~states/eV). The conduction-band minimum approaches $E_F$ near $M$.}
        \label{fig:suppl_tas2_2htheta}
    \end{minipage}
\end{figure}

\begin{figure}[H]
    \begin{minipage}{0.48\textwidth}
        \includegraphics[width=\linewidth]{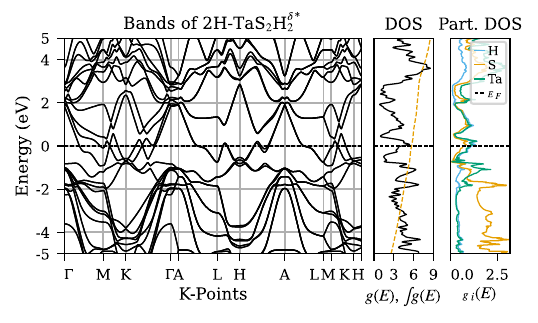}
        \caption{Band structure of 2H-TaS$_2$--H$_2^{\delta*}$. Metallic ($g(E_F)=4.43$~states/eV). The dispersion remains strongly hybridized along $\Gamma$--$A$.}
        \label{fig:suppl_tas2_h4deltaasterix}
    \end{minipage}\hfill
    \begin{minipage}{0.48\textwidth}
        \includegraphics[width=\linewidth]{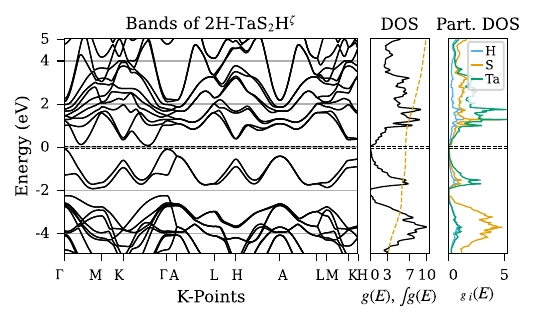}
        \caption{Band structure of 2H-TaS$_2$--H$^{\zeta}$. Semiconducting ($E_g=0.22$~eV). The small gap renders this structure susceptible to semimetallic behavior under small perturbations.}
        \label{fig:suppl_tas2_2hzeta}
    \end{minipage}
\end{figure}

\begin{figure}[H]
    \begin{minipage}{0.48\textwidth}
        \includegraphics[width=\linewidth]{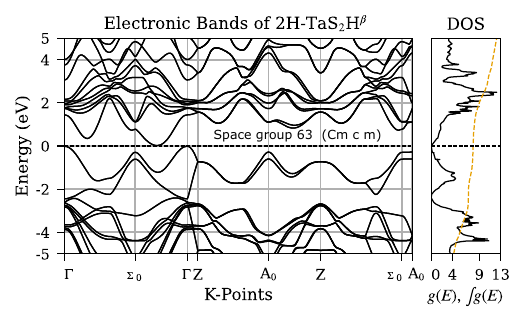}
        \caption{Band structure of 2H-TaS$_2$--H$^{\beta}$. Semiconducting ($E_g=0.06$~eV). The small gap forms at the Fermi level upon double $\beta$-site intercalation.}
        \label{fig:suppl_tas2_2hbeta}
    \end{minipage}\hfill
    \begin{minipage}{0.48\textwidth}
        \includegraphics[width=\linewidth]{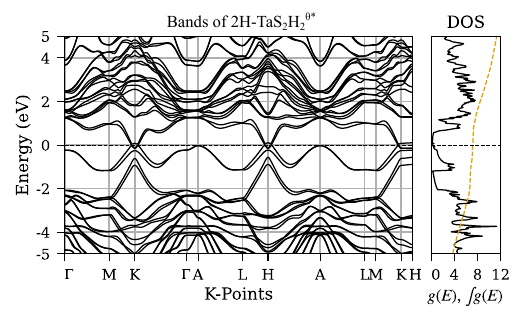}
        \caption{Band structure of 2H-TaS$_2$--H$_2^{\theta*}$. Metallic ($g(E_F)=0.10$~states/eV). The small residual DOS reflects minor band overlap near $E_F$.}
        \label{fig:suppl_tas2_2hthetamod}
    \end{minipage}
\end{figure}

\begin{figure}[H]
    \begin{minipage}{0.48\textwidth}
        \includegraphics[width=\linewidth]{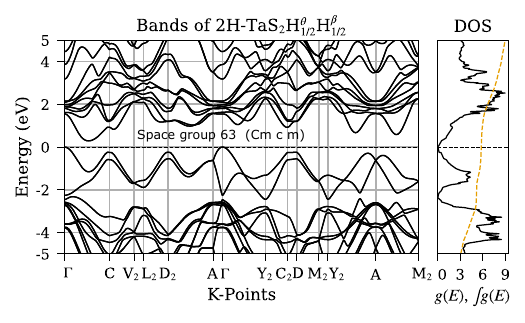}
        \caption{Band structure of 2H-TaS$_2$--H$_{1/2}^{\theta}$H$_{1/2}^{\beta}$. Metallic ($g(E_F)=0.01$~states/eV). Nearly gapped with very small DOS at $E_F$.}
        \label{fig:suppl_tas2_2hthetabeta}
    \end{minipage}\hfill
    \begin{minipage}{0.48\textwidth}
        \includegraphics[width=\linewidth]{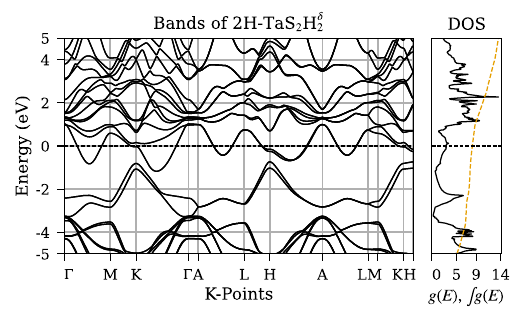}
        \caption{Band structure of 2H-TaS$_2$--H$_2^{\delta}$. Metallic ($g(E_F)=2.83$~states/eV). Multiple H-derived bands cross $E_F$ at this quadruple intercalation.}
        \label{fig:suppl_tas2_h4delta}
    \end{minipage}
\end{figure}

\begin{figure}[H]
    \centering
    \begin{minipage}{0.48\textwidth}
        \includegraphics[width=\linewidth]{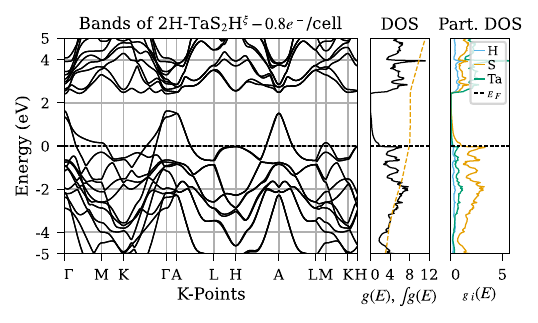}
        \caption{Band structure of 2H-TaS$_2$--H$^{\xi}$ under hole doping ($0.8e^-/\text{cell}$ removed). Metallic ($g(E_F)=1.94$~states/eV). The Fermi level descends into the valence bands, restoring metallicity.}
        \label{fig:suppl_tas2_xi_elec}
    \end{minipage}
\end{figure}

\subsection{\texorpdfstring{NbSe$_2$}{NbSe2}}

\begin{figure}[H]
    \centering
    \begin{minipage}{0.48\textwidth}
        \includegraphics[width=\linewidth]{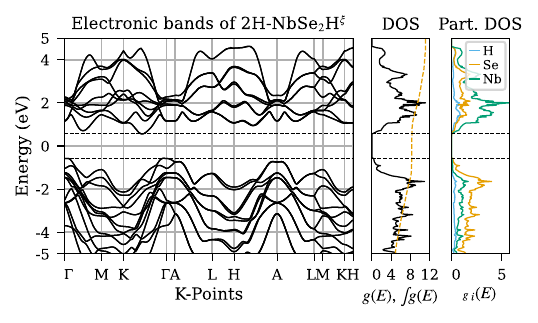}
        \caption{Band structure of 2H-NbSe$_2$--H$^{\xi}$. Semiconducting ($E_g=1.15$~eV). The gap opens cleanly across the Brillouin zone.}
        \label{fig:suppl_nbse2_2hxi}
    \end{minipage}
\end{figure}

\begin{figure}[H]
    \begin{minipage}{0.48\textwidth}
        \includegraphics[width=\linewidth]{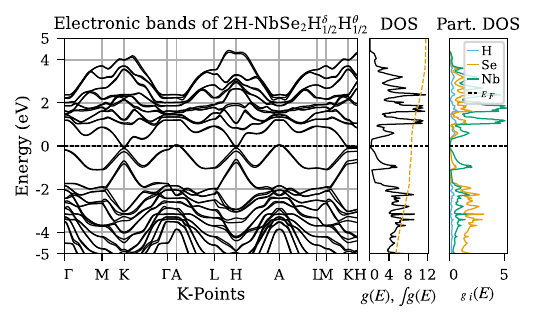}
        \caption{Band structure of 2H-NbSe$_2$--H$^{\delta/\theta}$. Metallic ($g(E_F)=0.18$~states/eV). The mixed-site configuration shows partial band hybridization near $E_F$.}
        \label{fig:suppl_nbse2_2hdeltatheta}
    \end{minipage}\hfill
    \begin{minipage}{0.48\textwidth}
        \includegraphics[width=\linewidth]{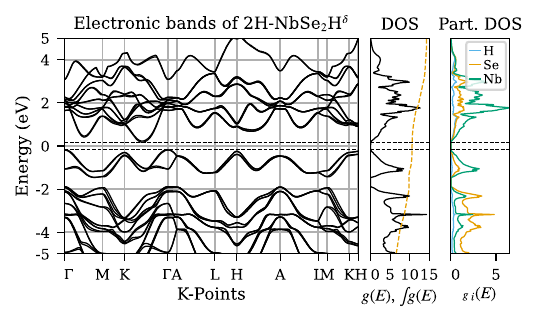}
        \caption{Band structure of 2H-NbSe$_2$--H$^{\delta}$. Semiconducting ($E_g=0.37$~eV).}
        \label{fig:suppl_nbse2_2hdelta}
    \end{minipage}
\end{figure}

\begin{figure}[H]
    \begin{minipage}{0.48\textwidth}
        \includegraphics[width=\linewidth]{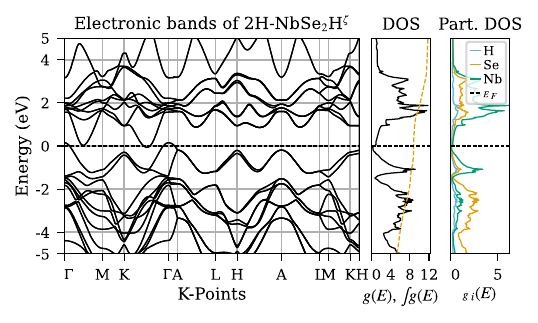}
        \caption{Band structure of 2H-NbSe$_2$--H$^{\zeta}$. Metallic ($g(E_F)=0.58$~states/eV). The intralayer $\zeta$ site preserves a small metallic overlap.}
        \label{fig:suppl_nbse2_2hzeta}
    \end{minipage}\hfill
    \begin{minipage}{0.48\textwidth}
        \includegraphics[width=\linewidth]{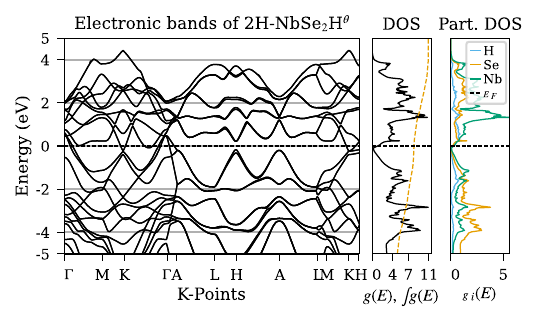}
        \caption{Band structure of 2H-NbSe$_2$--H$^{\theta}$. Metallic ($g(E_F)=1.00$~states/eV). The band edges touch $E_F$ near $K$.}
        \label{fig:suppl_nbse2_2htheta}
    \end{minipage}
\end{figure}

\clearpage
\section{Effect of $c$-axis compression on the electronic structure of H-intercalated \texorpdfstring{2H-TaS$_2$}{2H-TaS2}}
\label{sec:pressure_tas2}

We assess whether $c$-axis compression metallizes the semiconducting H$_2^{\xi}$ and H$_2^{\delta}$ phases. This perturbation primarily changes interlayer hybridization, so it directly tests whether the gap is tied to the expanded vdW spacing of the intercalated structures. For both, $E_F$ remains within the gap up to 10~GPa, ruling out mechanical compression as a viable metallization route.

\begin{figure}[H]
    \begin{minipage}{0.48\textwidth}
        \includegraphics[width=\linewidth]{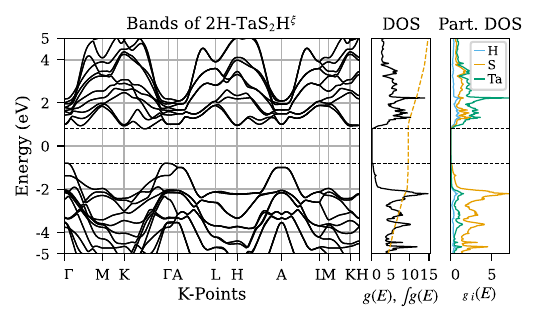}
        \caption{Band structure of 2H-TaS$_2$--H$^{\xi}$ at 0~GPa. Semiconducting ($E_g=1.60$~eV).}
        \label{fig:suppl_tas2_xi_0gpa}
    \end{minipage}\hfill
    \begin{minipage}{0.48\textwidth}
        \includegraphics[width=\linewidth]{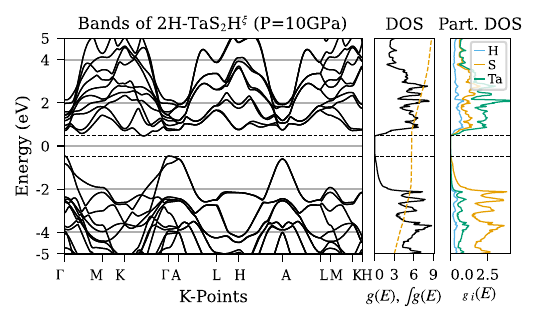}
        \caption{Band structure of 2H-TaS$_2$--H$^{\xi}$ at 10~GPa. Semiconducting ($E_g=0.96$~eV).}
        \label{fig:suppl_tas2_xi_10gpa}
    \end{minipage}
\end{figure}

\begin{figure}[H]
    \begin{minipage}{0.48\textwidth}
        \includegraphics[width=\linewidth]{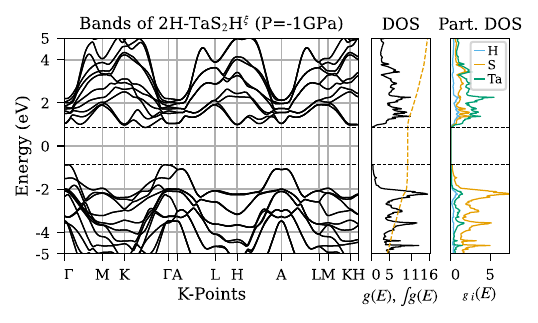}
        \caption{Band structure of 2H-TaS$_2$--H$^{\xi}$ at $-1$~GPa. Semiconducting ($E_g=1.75$~eV).}
        \label{fig:suppl_tas2_xi_m1gpa}
    \end{minipage}\hfill
    \begin{minipage}{0.48\textwidth}
        \includegraphics[width=\linewidth]{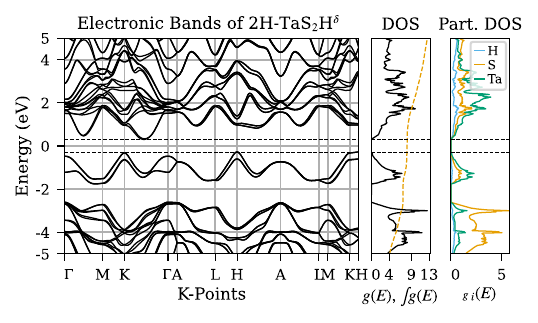}
        \caption{Band structure of 2H-TaS$_2$--H$^{\delta}$ at 0~GPa. Semiconducting ($E_g=0.62$~eV).}
        \label{fig:suppl_tas2_delta_0gpa}
    \end{minipage}
\end{figure}

\begin{figure}[H]
    \begin{minipage}{0.48\textwidth}
        \includegraphics[width=\linewidth]{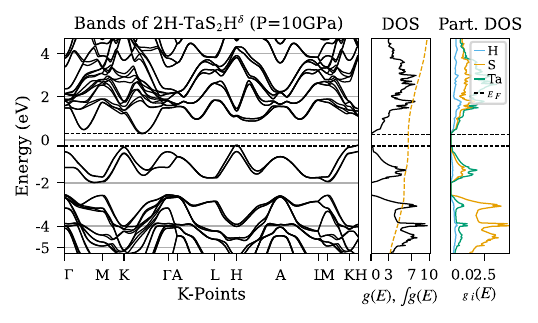}
        \caption{Band structure of 2H-TaS$_2$--H$^{\delta}$ at 10~GPa. Semiconducting ($E_g=0.58$~eV).}
        \label{fig:suppl_tas2_delta_10gpa}
    \end{minipage}\hfill
    \begin{minipage}{0.48\textwidth}
        \includegraphics[width=\linewidth]{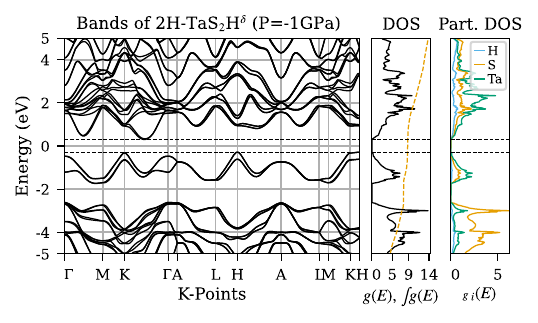}
        \caption{Band structure of 2H-TaS$_2$--H$^{\delta}$ at $-1$~GPa. Semiconducting ($E_g=0.62$~eV).}
        \label{fig:suppl_tas2_delta_m1gpa}
    \end{minipage}
\end{figure}

$c$-axis compression narrows the gap (H$^{\xi}$: $1.60 \rightarrow 0.96$~eV over $0$--$10$~GPa, and H$^{\delta}$: $0.62$~eV at both $0$~GPa and $-1$~GPa), but $E_F$ remains within the gap throughout and metallicity is not restored. The persistence of the gap shows that the semiconducting state is not merely a consequence of excess layer separation but more a consequence of the band filling discussed in the main text.

\section{Effect of in-plane biaxial strain on the electronic structure of H-intercalated \texorpdfstring{2H-TaS$_2$}{2H-TaS2}}
\label{sec:strain_tas2}

In-plane biaxial strain—which modifies the lateral lattice constant with $c$ free to relax—is evaluated as an alternative metallization route for H$^{\xi}$ and H$^{\delta}$. Unlike $c$-axis pressure, this perturbation changes intralayer hybridization more directly and therefore probes whether the gapped phases lie close to an in-plane band-overlap transition. Strain is an obvious control parameter in 2H-TMDs, particularly due to their vdW bonding: recent work on 2H-NbSe$_2$ has shown that both the CDW response and the superconducting transition can be strongly reshaped by lattice deformation~\cite{Kundu_2024,HenriquezGuerra_2024}.

\begin{figure}[H]
    \begin{minipage}{0.48\textwidth}
        \includegraphics[width=\linewidth]{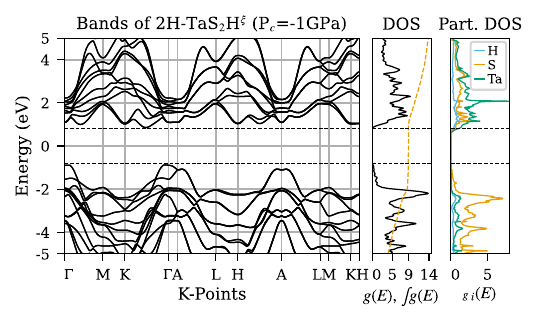}
        \caption{Band structure of 2H-TaS$_2$--H$^{\xi}$ under $-1$~GPa biaxial strain. Semiconducting ($E_g=1.76$~eV). Tensile strain slightly enlarges the gap.}
        \label{fig:suppl_tas2_xi_strain_m1gpa}
    \end{minipage}\hfill
    \begin{minipage}{0.48\textwidth}
        \includegraphics[width=\linewidth]{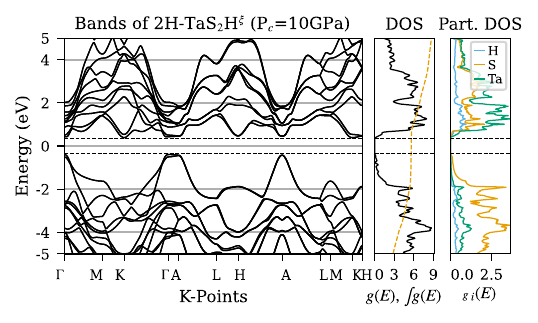}
        \caption{Band structure of 2H-TaS$_2$--H$^{\xi}$ under $+10$~GPa biaxial strain. Semiconducting ($E_g=0.82$~eV). Strong compressive strain narrows the gap significantly.}
        \label{fig:suppl_tas2_xi_strain_10gpa}
    \end{minipage}
\end{figure}

\begin{figure}[H]
    \begin{minipage}{0.48\textwidth}
        \includegraphics[width=\linewidth]{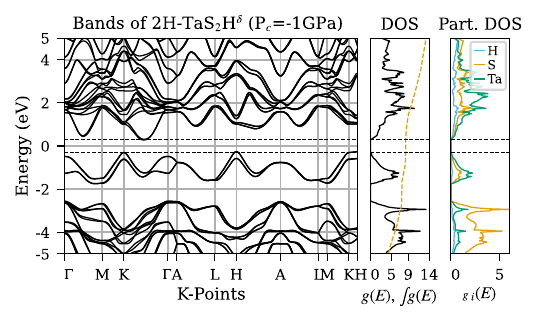}
        \caption{Band structure of 2H-TaS$_2$--H$^{\delta}$ under $-1$~GPa biaxial strain. Semiconducting ($E_g=0.58$~eV). The gap is marginally reduced relative to the unstrained case.}
        \label{fig:suppl_tas2_delta_strain_m1gpa}
    \end{minipage}\hfill
    \begin{minipage}{0.48\textwidth}
        \includegraphics[width=\linewidth]{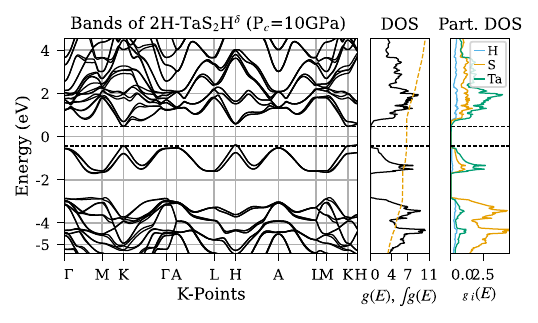}
        \caption{Band structure of 2H-TaS$_2$--H$^{\delta}$ under $+10$~GPa biaxial strain. Metallic ($g(E_F)=0.14$~states/eV). Strong in-plane compression restores metallicity in the $\delta$ phase.}
        \label{fig:suppl_tas2_delta_strain_10gpa}
    \end{minipage}
\end{figure}

For H$^{\xi}$, the gap narrows from $1.76$~eV (tensile) to $0.82$~eV ($+10$~GPa), with $E_F$ remaining within the gap. For H$^{\delta}$, $+10$~GPa compressive strain restores metallicity ($g(E_F)=0.14$~states/eV), though the required magnitude is experimentally challenging. The contrasting response shows that H$^{\delta}$ lies much closer to a band-overlap threshold than H$^{\xi}$, even though both are semiconducting at equilibrium.

\clearpage
\section{Various phonon band structures}

Phonon dispersions and partial PhDOS for configurations not shown in the main text are presented here, organized by material. Across the full set, two motifs are most informative: the fate of the CDW soft mode near $\mathbf{q}_\mathrm{CDW}\approx\tfrac{2}{3}\overline{\Gamma\mathrm{M}}$ and the appearance of isolated high-frequency H-derived optical branches above the host phonon spectrum. Reading the figures in this order makes it clear which structures are stabilized by gap opening and which simply relocate the instability.

\subsection{\texorpdfstring{TaS$_2$}{TaS2}}

The TaS$_2$ configurations fall into three groups. H$^{\xi}$, H$^{\delta}$, and H$^{\zeta}$ are fully stabilized and gapped. The remaining double-intercalated structures retain part of the original CDW character, and the quadruple-intercalated structures develop new imaginary modes once H--H repulsion becomes dominant.

\begin{figure}[H]
    \begin{minipage}{0.48\textwidth}
        \includegraphics[width=\linewidth]{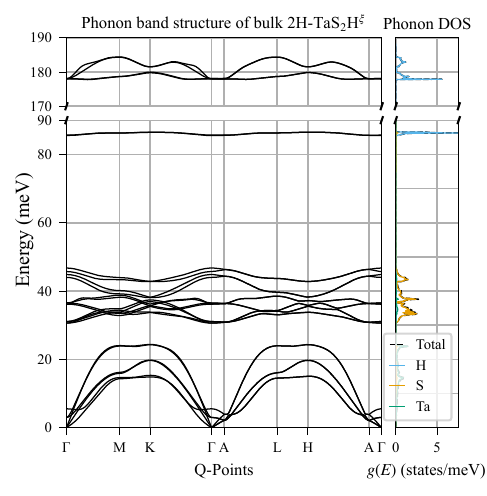}
        \caption{Phonon dispersion and PhDOS of 2H-TaS$_2$--H$^{\xi}$. The CDW soft mode is fully suppressed and the spectrum is dynamically stable. An isolated H-derived optical branch appears near $\sim\!179$~meV, well above the host-lattice phonon spectrum. This is the highest-binding intralayer configuration.}
        \label{fig:suppl_phdos_TaS2H2xi}
    \end{minipage}\hfill
    \begin{minipage}{0.48\textwidth}
        \includegraphics[width=\linewidth]{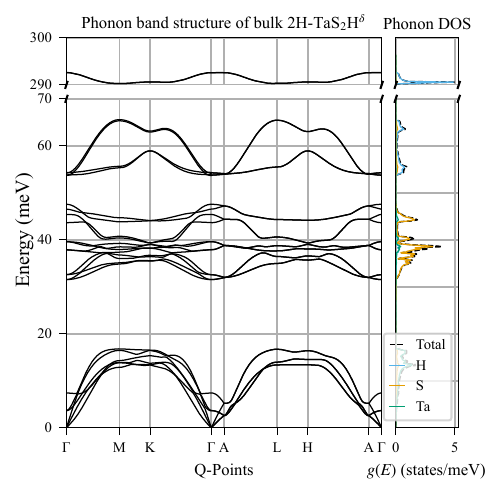}
        \caption{Phonon dispersion and PhDOS of 2H-TaS$_2$--H$^{\delta}$. CDW fully suppressed, dynamically stable. The H-derived optical branch is shifted to $\sim\!290$~meV, higher than in H$^{\xi}$.}
        \label{fig:suppl_phdos_TaS2H2delta}
    \end{minipage}
\end{figure}

\begin{figure}[H]
    \begin{minipage}{0.48\textwidth}
        \includegraphics[width=\linewidth]{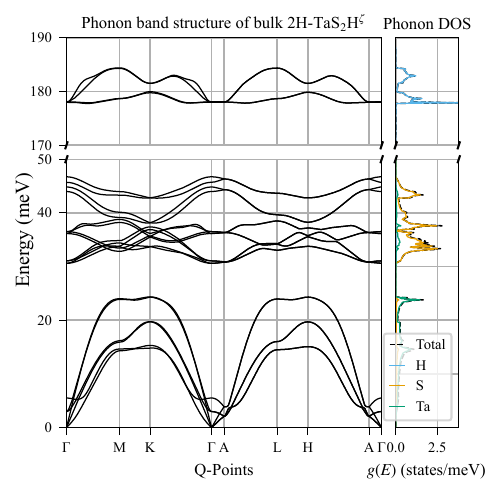}
        \caption{Phonon dispersion and PhDOS of 2H-TaS$_2$--H$^{\zeta}$. CDW suppressed, dynamically stable.}
        \label{fig:suppl_phdos_TaS2H2zeta}
    \end{minipage}\hfill
    \begin{minipage}{0.48\textwidth}
        \includegraphics[width=\linewidth]{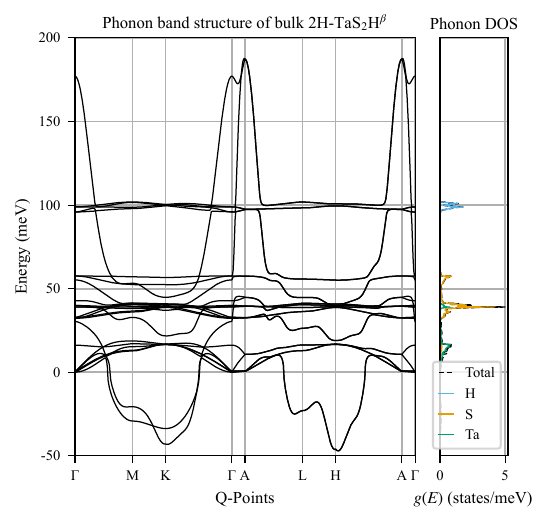}
        \caption{Phonon dispersion and PhDOS of 2H-TaS$_2$--H$^{\beta}$. Note that the k-path used does not represent the proper path for this structure, which has space group 63.}
        \label{fig:suppl_phdos_TaS2H2beta}
    \end{minipage}
\end{figure}

\begin{figure}[H]
    \begin{minipage}{0.48\textwidth}
        \includegraphics[width=\linewidth]{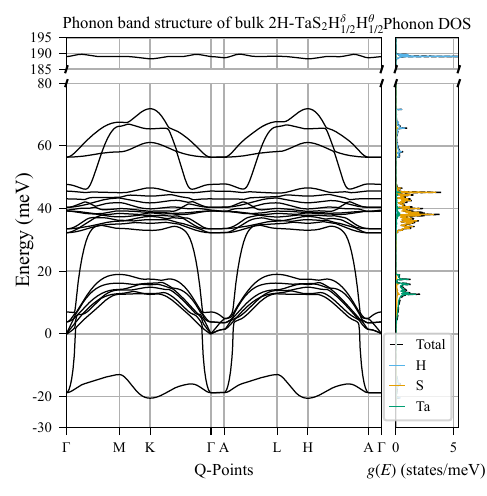}
        \caption{Phonon dispersion and PhDOS of 2H-TaS$_2$--H$^{\delta/\theta}$ (mixed site). The CDW soft mode persists, with a fully dynamically unstable branch at $-20$~meV.}
        \label{fig:suppl_phdos_TaS2H2delta_theta}
    \end{minipage}\hfill
    \begin{minipage}{0.48\textwidth}
        \includegraphics[width=\linewidth]{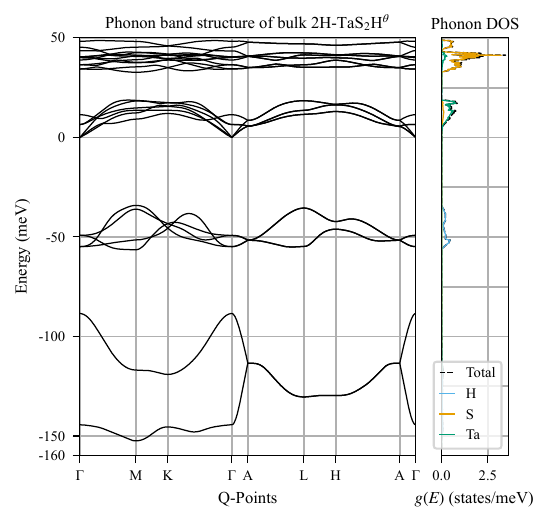}
        \caption{Phonon dispersion and PhDOS of 2H-TaS$_2$--H$^{\theta}$. The CDW soft mode near $\mathbf{q}_\mathrm{CDW}$ is not lifted, and has instead become fully imaginary alongside four bands dominated by hydrogen near $-50$~meV.}
        \label{fig:suppl_phdos_TaS2H2theta}
    \end{minipage}
\end{figure}

\begin{figure}[H]
    \begin{minipage}{0.48\textwidth}
        \includegraphics[width=\linewidth]{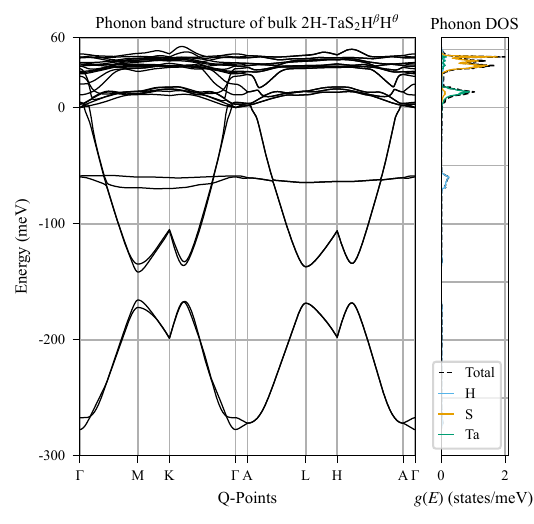}
        \caption{Phonon dispersion and PhDOS of 2H-TaS$_2$--H$_2^{\beta/\theta}$ (variant). Completely dynamically unstable with a reduction of the space group (63).}
        \label{fig:suppl_phdos_TaS2H2beta_2theta}
    \end{minipage}\hfill
    \begin{minipage}{0.48\textwidth}
        \includegraphics[width=\linewidth]{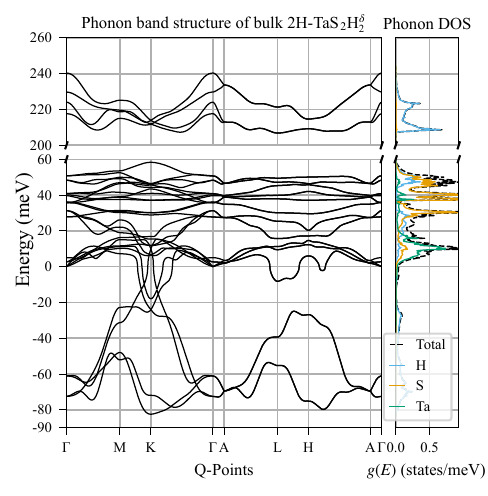}
        \caption{Phonon dispersion and PhDOS of 2H-TaS$_2$--H$_2^{\delta}$. Imaginary modes re-emerge, including near $\mathbf{q}=K$, together with fully imaginary modes up to $-80$~meV.}
        \label{fig:suppl_phdos_TaS2H4delta}
    \end{minipage}
\end{figure}

\begin{figure}[H]
    \centering
    \begin{minipage}{0.48\textwidth}
        \includegraphics[width=\linewidth]{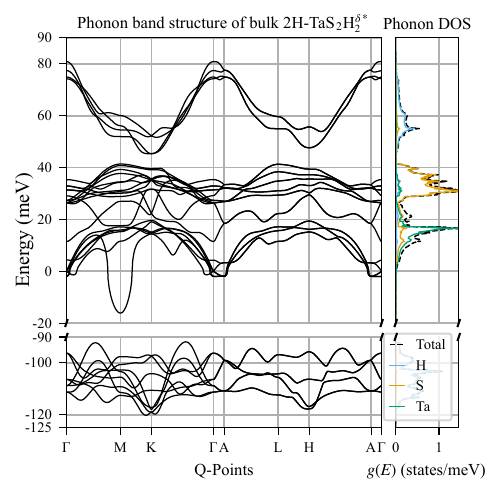}
        \caption{Phonon dispersion and PhDOS of 2H-TaS$_2$--H$_2^{\delta^*}$. Additional imaginary modes indicate a fully dynamically unstable structure, more severe than H$_2^{\delta}$.}
        \label{fig:suppl_phdos_TaS2H4deltaasterix}
    \end{minipage}
\end{figure}

\subsection{\texorpdfstring{NbSe$_2$}{NbSe2}}

NbSe$_2$ follows the same hierarchy as TaS$_2$, but with a weaker underlying CDW tendency. As a result, the stable phases are cleaner apart from the residual imaginary modes near $\Gamma$, of the order of $1$~meV, while the unstable mixed or quadruple phases still show the same re-emergence of CDW-related softening. The CDW soft mode in pristine NbSe$_2$ is weaker than in TaS$_2$.

\begin{figure}[H]
    \begin{minipage}{0.48\textwidth}
        \includegraphics[width=\linewidth]{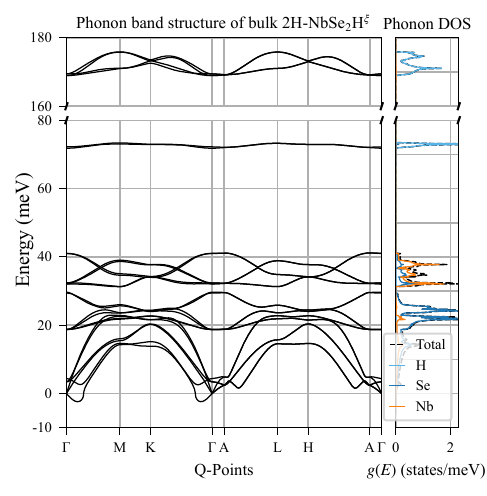}
        \caption{Phonon dispersion and PhDOS of 2H-NbSe$_2$--H$^{\xi}$. CDW fully suppressed, dynamically stable. The large electronic gap ($E_g\approx1.15$~eV) strongly stiffens the low-frequency modes. An isolated H-derived optical branch is visible at high frequency. Only residual imaginary frequencies remain at $\Gamma$.}
        \label{fig:suppl_phdos_NbSe2H2xi}
    \end{minipage}\hfill
    \begin{minipage}{0.48\textwidth}
        \includegraphics[width=\linewidth]{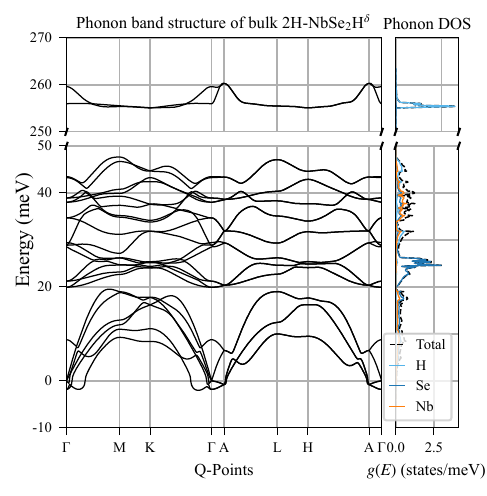}
        \caption{Phonon dispersion and PhDOS of 2H-NbSe$_2$--H$^{\delta}$. CDW suppressed, dynamically stable with residual imaginary modes near $\Gamma$.}
        \label{fig:suppl_phdos_NbSe2H2delta}
    \end{minipage}
\end{figure}

\begin{figure}[H]
    \begin{minipage}{0.48\textwidth}
        \includegraphics[width=\linewidth]{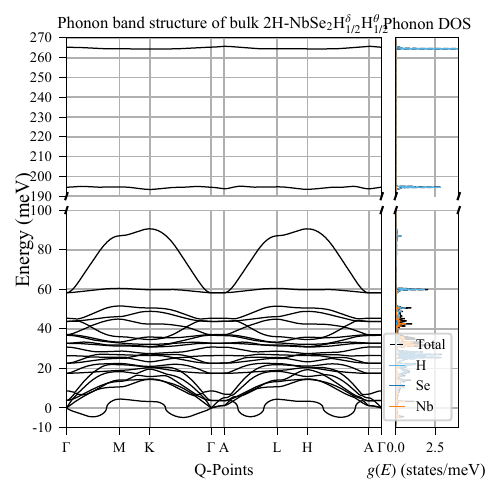}
        \caption{Phonon dispersion and PhDOS of 2H-NbSe$_2$--H$^{\delta/\theta}$. Residual CDW softening near $\mathbf{q}_\mathrm{CDW}$ persists.}
        \label{fig:suppl_phdos_NbSe2H2delta_theta}
    \end{minipage}\hfill
    \begin{minipage}{0.48\textwidth}
        \includegraphics[width=\linewidth]{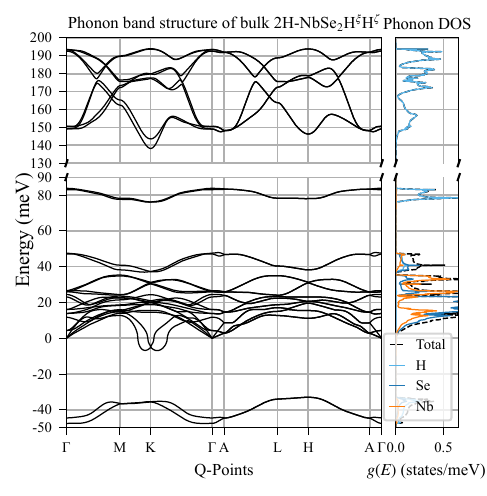}
        \caption{Phonon dispersion and PhDOS of 2H-NbSe$_2$--H$_2^{\xi/\zeta}$ ($N_H=4$). CDW imaginary modes re-emerge upon quadruple intercalation, mirroring the TaS$_2$ trend.}
        \label{fig:suppl_phdos_NbSe2H4xi_zeta}
    \end{minipage}
\end{figure}
\clearpage
\section{Effect on phonon band structure of van der Waals and Spin-Orbit coupling}

Four reference calculations for pristine 2H-TaS$_2$ with vdW and SOC selectively disabled justify their inclusion in all DFPT calculations.

\begin{figure}[H]
    \begin{minipage}{0.48\textwidth}
        \includegraphics[width=\linewidth]{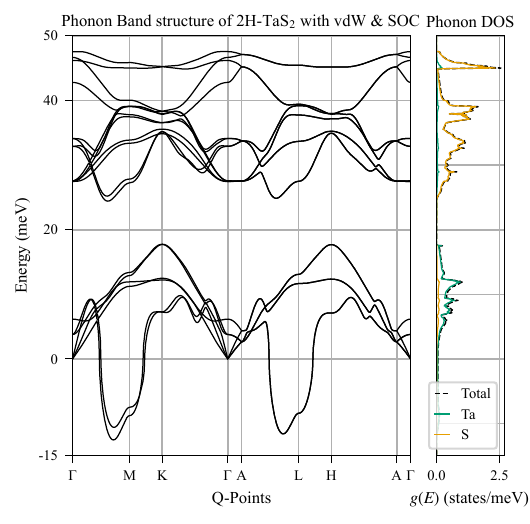}
        \caption{Phonon band structure of pristine 2H-TaS$_2$ with van der Waals and spin-orbit interactions enabled (reference calculation). A single CDW soft mode appears near $\mathbf{q}_\mathrm{CDW}$, in agreement with experiment.}
        \label{fig:suppl_tas2_vdw_so}
    \end{minipage}\hfill
    \begin{minipage}{0.48\textwidth}
        \includegraphics[width=\linewidth]{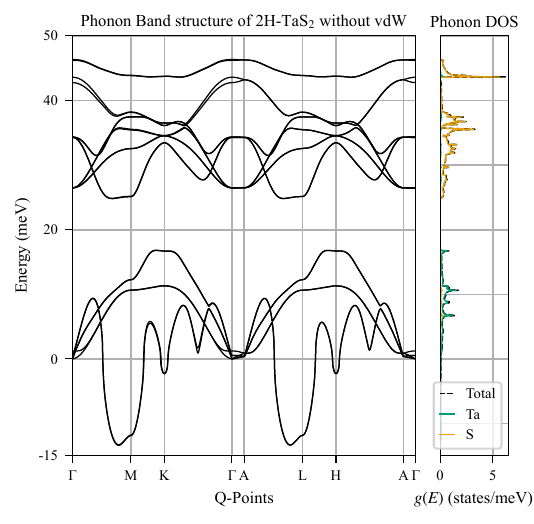}
        \caption{Phonon band structure of pristine 2H-TaS$_2$ without van der Waals interactions (SOC enabled). A spurious imaginary mode appears at $K$ due to the absence of long-range interlayer cohesion.}
        \label{fig:suppl_tas2_no_vdw}
    \end{minipage}
    \begin{minipage}{0.48\textwidth}
        \includegraphics[width=\linewidth]{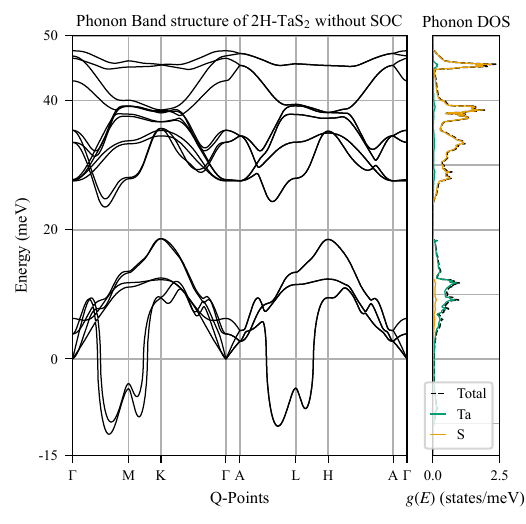}
        \caption{Phonon band structure of pristine 2H-TaS$_2$ without spin-orbit coupling (vdW enabled). The CDW soft mode splits into two distinct minima near $\mathbf{q}_\mathrm{CDW}$, reflecting the absence of relativistic splitting of the Ta $5d$ states.}
        \label{fig:suppl_tas2_no_so}
    \end{minipage}\hfill
    \begin{minipage}{0.48\textwidth}
        \includegraphics[width=\linewidth]{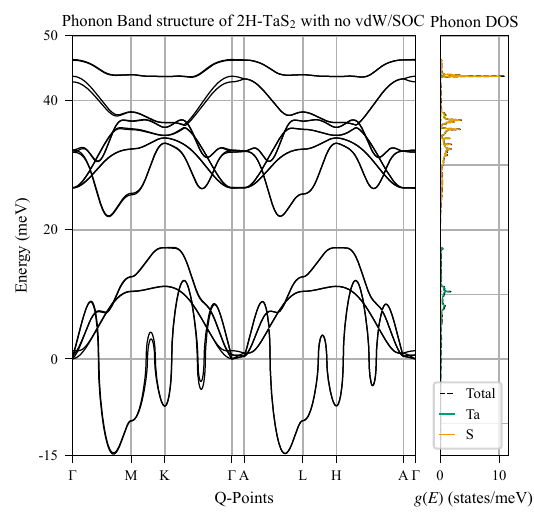}
        \caption{Phonon band structure of pristine 2H-TaS$_2$ with neither van der Waals nor spin-orbit interactions. Both artefacts compound: the CDW minimum is split and additional imaginary modes appear, yielding a qualitatively incorrect spectrum.}
        \label{fig:suppl_tas2_no_vdw_so}
    \end{minipage}
\end{figure}

Without vdW corrections (Fig.~\ref{fig:suppl_tas2_no_vdw}), an imaginary mode appears at $K$, absent experimentally. Without SOC (Fig.~\ref{fig:suppl_tas2_no_so}), the single CDW minimum near $\mathbf{q}_\mathrm{CDW}$ splits into a double well, inconsistent with the observed $3\times3$ superstructure. Disabling both (Fig.~\ref{fig:suppl_tas2_no_vdw_so}) compounds both artefacts. Only the full calculation (Fig.~\ref{fig:suppl_tas2_vdw_so}) reproduces the single CDW soft mode at the correct wavevector.

\clearpage

\section{Isotope effect}
To further assess which atomic vibrations control the superconducting instability, we examined the isotope dependence of the calculated critical temperature $T_c$ for hole-doped H$^{\xi}$-TaS$_2$ ($-0.8e^-$/cell, as in the main text) by changing the mass of each atomic species separately while keeping the electronic structure and the force constants otherwise unchanged. These calculations used a larger smearing than the 2.5~mHa of the main text, so the absolute $T_c$ ($\approx 0.16$~K at natural masses) differs from the value quoted there (0.071~K). Only the mass dependence is used here. For each element $X=\mathrm{H},\mathrm{S},\mathrm{Ta}$, the resulting $T_c$ values were fitted to:
\begin{equation}
T_c(X) = C_X M_X^{-\alpha_X},
\label{eq:isotope_scaling}
\end{equation}
where $M_X$ is the isotope mass obtained from the Nubase database~\cite{AUDI20033}, $C_X$ is a prefactor, and $\alpha_X$ is the isotope exponent. The masses considered for each element are displayed in Table~\ref{tab:isotope_masses}.

\begin{table}[H]
  \centering
    \begin{tabular}{llllll}
    \toprule
    Isotope & Mass (u) & Isotope & Mass (u) & Isotope & Mass (u) \\
    \midrule
    \midrule
    H$^1$ & 1.0078 & S$^{32}$ & 31.972 & Ta$^{179}$ & 178.95 \\
    H$^2$ & 2.0141 & S$^{33}$ & 32.971 & Ta$^{180}$ & 179.95 \\
    H$^3$ & 3.0160 & S$^{34}$ & 33.968 & Ta$^{181}$ & 180.95 \\
    H$^4$ & 4.0264 & S$^{35}$ & 34.969 & Ta$^{182}$ & 181.95 \\
    H$^5$ & 5.0353 & S$^{36}$ & 35.967 & Ta$^{183}$ & 182.95 \\
    \bottomrule
    \end{tabular}%
    \caption{Masses for each isotope labeled by mass number.}
    \label{tab:isotope_masses}%
\end{table}%

\begin{figure}[H]
    \begin{minipage}{0.48\textwidth}
        \includegraphics[width=\linewidth]{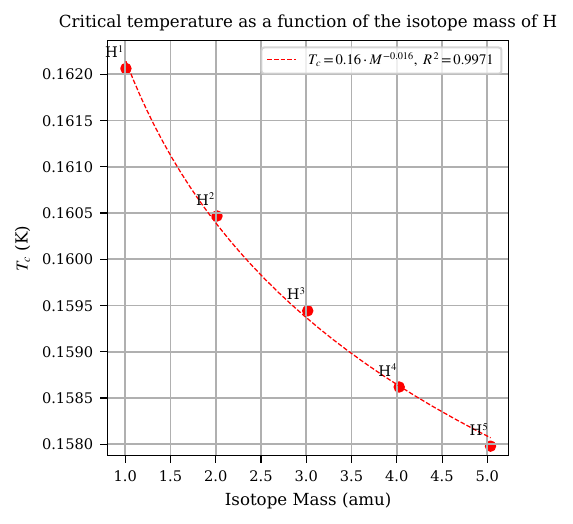}
        \caption{Calculated critical temperature $T_c$ as a function of hydrogen isotope mass, together with a fit to $T_c = C M^{-\alpha}$.}
        \label{fig:suppl_tas2_h2xi_isotope}
    \end{minipage}\hfill
    \begin{minipage}{0.48\textwidth}
        \includegraphics[width=\linewidth]{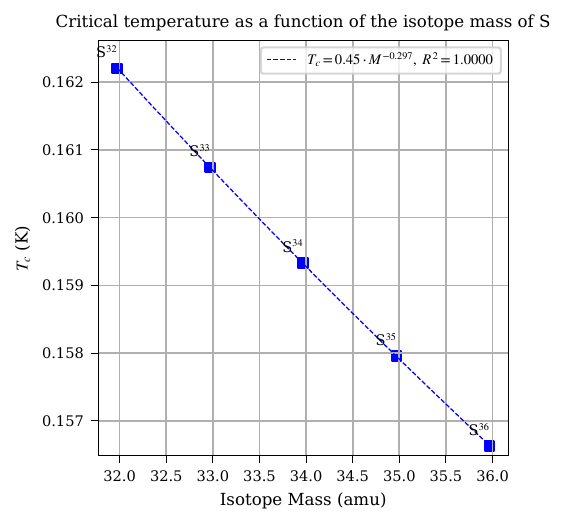}
        \caption{Calculated critical temperature $T_c$ as a function of sulfur isotope mass, together with a fit to $T_c = C M^{-\alpha}$.}
        \label{fig:suppl_tas2_h2delta_isotope}
    \end{minipage}
\end{figure}

\begin{figure}[H]
    \centering
    \begin{minipage}{0.48\textwidth}
        \includegraphics[width=\linewidth]{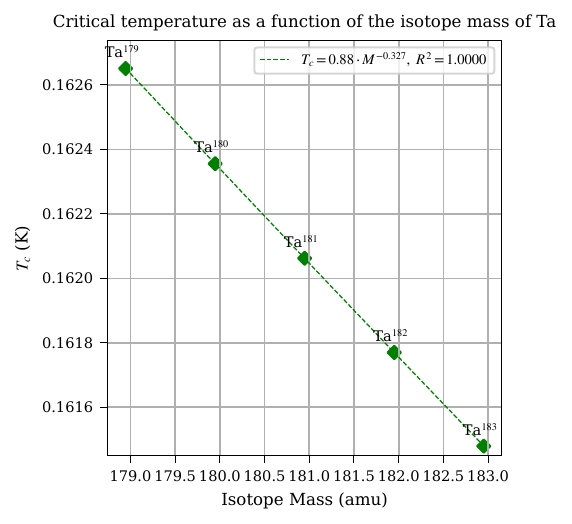}
        \caption{Calculated critical temperature $T_c$ as a function of tantalum isotope mass, together with a fit to $T_c = C M^{-\alpha}$.}
        \label{fig:suppl_tas2_h2zeta_isotope}
    \end{minipage}
\end{figure}

The calculated $T_c(M)$ curves are shown in Figs.~\ref{fig:suppl_tas2_h2xi_isotope}--\ref{fig:suppl_tas2_h2zeta_isotope}. In all three cases, $T_c$ decreases monotonically with increasing isotope mass, consistent with the expected mass renormalization of the phonon frequencies. The extracted isotope exponents are small for H, $\alpha_{\mathrm H}\approx 0.016$, and substantially larger for S and Ta, $\alpha_{\mathrm S}\approx 0.297$ and $\alpha_{\mathrm{Ta}}\approx 0.327$. The fits are excellent over the full mass range considered, with $R^2=0.9971$ for H and essentially unity for S and Ta.

These results show that the isotope response is dominated by the heavier host-lattice atoms rather than by hydrogen. In particular, the very small hydrogen isotope exponent indicates that the H-derived vibrations make only a minor contribution to the total pairing strength in this structure. This is consistent with the phonon spectra and Eliashberg function discussed in the main text, where the high-frequency H modes carry comparatively little weight in the electron-phonon coupling. By contrast, the larger exponents for S and Ta identify the host-lattice vibrations as the primary contributors to the superconductivity.

Overall, the isotope analysis supports the same physical picture inferred from the mode-resolved electron-phonon properties: hydrogen mainly modifies the system indirectly through structural and electronic effects, while the dominant pairing contribution originates from Ta- and S-related phonon modes.

\clearpage

\section{Convergence parameters}\label{sec:convergence}

Numerical convergence settings are summarized in Table~\ref{tab:convergence}.

\begin{table}[H]
    \caption{Convergence parameters used throughout the calculations.}
    \label{tab:convergence}
    \begin{ruledtabular}
    \begin{tabular}{lccc}
        Parameter & Values tested & Converged value & Criterion \\ \hline
        Plane-wave cutoff (Ha) & 20, 25, 30, 35, 40, 45, 50, 55, 60, 65, 70, 75, 80, 85 & 50 & $\Delta E < 1$~meV/atom \\
        Electronic $k$-mesh & $3\times3\times1$, $6\times6\times2$, $\ldots$, $33\times33\times11$ & $16\times16\times6$ & $\Delta E < 0.1$~meV/atom \\
        Phonon $q$-mesh & $k$-mesh$/2$ & $8\times8\times3$ & See $k$-mesh and cutoff \\
        Occupation smearing (mHa) & 2.5, 5.0, 7.5, 10.0 & 2.5 (=68.0~meV) & $\Delta\omega_{\mathrm{ph}} < 1$~meV \\
        Wavefunction tolerance & $10^{-20}$--$10^{-24}$ & $<10^{-24}$ & Convention \\
        Potential tolerance & $10^{-10}$--$10^{-12}$ & $<10^{-12}$ & Convention \\
    \end{tabular}
    \end{ruledtabular}
\end{table}

\begin{table}[H]
    \caption{$\Gamma$-point phonon frequencies of H$^{\delta}$-TaS$_2$ at plane-wave cutoffs of $50$ and $80$~Ha, before the acoustic sum rule is enforced. Modes are sorted by frequency and negative values denote imaginary frequencies.}
    \label{tab:gamma_cutoff}
    \begin{ruledtabular}
    \begin{tabular}{lccc}
        Modes & $\omega_{50}$ (meV) & $\omega_{80}$ (meV) & $\omega_{80}-\omega_{50}$ (meV) \\ \hline
        1--2 (acoustic) & $-1.064$ & $-0.893$ & $+0.171$ \\
        3 (acoustic) & $-0.032$ & $+0.081$ & $+0.113$ \\
        4--5 & $0.193$ & $0.141$ & $-0.052$ \\
        6 & $7.822$ & $7.820$ & $-0.001$ \\
        7--22 & $30.79$--$51.64$ & $30.84$--$51.81$ & $|\Delta|\leq0.162$ \\
        23 (H stretch) & $291.41$ & $291.82$ & $+0.410$ \\
        24 (H stretch) & $294.95$ & $295.33$ & $+0.378$ \\
    \end{tabular}
    \end{ruledtabular}
\end{table}

In the NbSe$_2$ structures, weakly imaginary modes persist near $\Gamma$ with magnitudes of the order of $1$~meV. To test whether they are limited by the plane-wave cutoff, the $\Gamma$-point phonon frequencies of H$^{\delta}$-TaS$_2$ were recomputed at $E_{\mathrm{cut}}=80$~Ha and compared with the $50$~Ha production values (Table~\ref{tab:gamma_cutoff}). All optical frequencies agree within $0.41$~meV ($3.3$~cm$^{-1}$), with the largest deviation ($0.14\%$) for the two H-stretching modes, and all other optical frequencies agree within $0.16$~meV ($1.3$~cm$^{-1}$). The relaxed lattice parameters agree within $0.001$~\AA\ and the binding energy changes by $12$~meV/H (2958 vs.\ 2946~meV/H, Table~\ref{tab:supp_binding} and main-text Table~I). In contrast, the residual frequency of the acoustic modes at $\Gamma$, which should vanish by the acoustic sum rule, is almost insensitive to the cutoff and only changes from $-1.06$ to $-0.89$~meV. These residuals are therefore only weakly reduced by a larger cutoff and are of the size expected for a small numerical violation of the acoustic sum rule ($\approx0.4\%$ of the highest phonon frequency). They are roughly $20$ to $80$ times smaller than the imaginary frequencies of the dynamically unstable structures (Figs.~\ref{fig:suppl_phdos_TaS2H2delta_theta}, \ref{fig:suppl_phdos_TaS2H2theta} and \ref{fig:suppl_phdos_TaS2H4delta}). Given their energy scale and localization near $\Gamma$, their contribution to $\lambda$ is negligible and does not affect the trends reported in the main text. Such tests are computationally prohibitive with the given resources for all structures, so $50$~Ha is used throughout. 

The plane-wave cutoff was determined by two criteria, total-energy convergence ($1$~meV/atom at $50$~Ha) and, more stringently, phonon frequency convergence. The $\Gamma$-point phonon frequencies of H$^{\delta}$-TaS$_2$ at $50$ and $80$~Ha agree within $3.3$~cm$^{-1}$ ($0.41$~meV) and most modes agree within $1$~cm$^{-1}$ (Table~\ref{tab:gamma_cutoff}), confirming convergence at $50$~Ha.

\begin{figure}[H]
    \centering
    \includegraphics[width=0.95\textwidth]{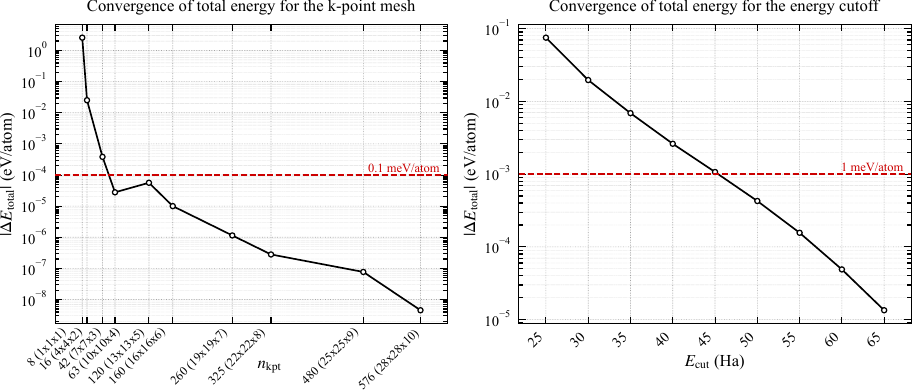}
    \caption{Total energy differences between successive values for 2H-TaS$_2$ (with vdW and SOC) for different k-point meshes and plane-wave energy cutoffs, with the convergence criteria denoted by the dotted line.}
    \label{fig:suppl_cutoff_convergence}
\end{figure}

\begin{figure}[H]
    \centering
    \includegraphics[width=0.95\textwidth]{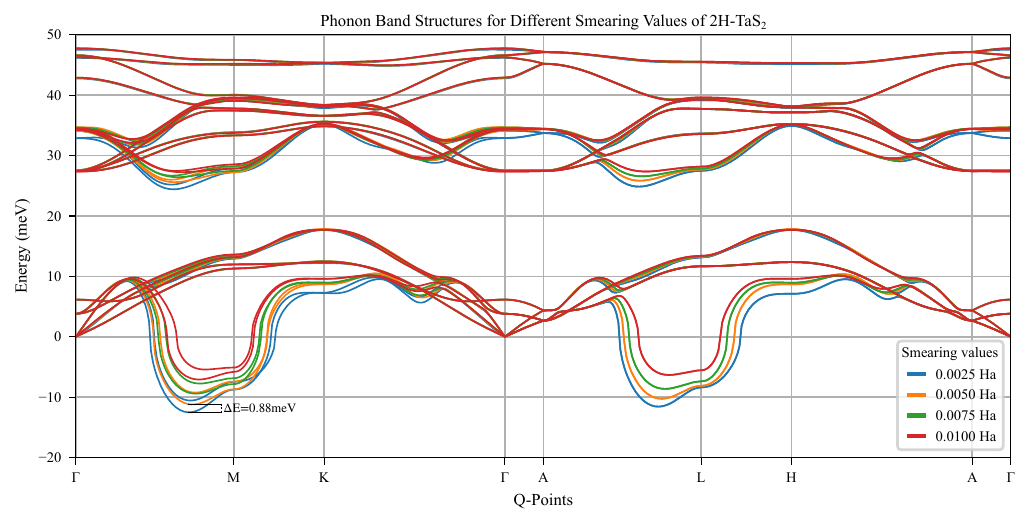}
    \caption{Phonon band structures for 2H-TaS$_2$ (with vdW and SOC) for different smearing values.}
    \label{fig:suppl_phonon_convergence}
\end{figure}

The electronic smearing was converged by requiring phonon frequencies at consecutive smearing values to agree within $1$~meV throughout the Brillouin zone. For pristine TaS$_2$, calculations at 2.5, 5.0, 7.5, and 10.0~mHa show a maximum frequency difference of $0.88$~meV between the 2.5 and 5.0~mHa cases (Fig.~\ref{fig:suppl_phonon_convergence}), confirming that $2.5$~mHa is sufficient.

\bibliographystyle{apsrev4-2}
\bibliography{references}